\documentclass{acmtog}

\usepackage{times}
\usepackage{epsfig}
\usepackage{graphicx}
\usepackage{amsmath}
\usepackage{amssymb}
\usepackage{enumerate}
\usepackage{bm}
\usepackage[normalem]{ulem}
\usepackage[usenames]{xcolor}

\acmVolume{35}
\acmNumber{1}
\acmYear{2015}
\acmMonth{December}
\acmArticleNum{4}  
\acmdoi{10.1145/2832905}

\graphicspath{{./}}

\title{Pushing the Limits of 3D Color Printing: \\Error Diffusion with Translucent Materials}

\author{Alan Brunton, Can Ates Arikan {\upshape and}  Philipp Urban
\affil{Fraunhofer Institute for Computer Graphics Research IGD}
}

\begin{document}

\terms{3D color printing, half-toning, error diffusion}

\maketitle

\begin{bottomstuff} 
contact: \{alan.brunton,can.ates.arikan,philipp.urban\}@igd.fraunhofer.de\\
web: \url{www.cuttlefish.de}
\end{bottomstuff}

\definecolor{checkmecol}{rgb}{0.7,0.2,0.3}
\newcommand{\todo}[1]{\PackageWarning{}{Unprocessed todo}{\textcolor{checkmecol}{\textbf{TODO} \textbf{#1}}}}
\newcommand{\anno}[1]{\PackageWarning{}{Unprocessed anno}{{\normalsize\textit{\textcolor{cyan}{\textbf{[Annotation:} #1\textbf{]}}}}}}

\newcommand{\qheading}[1]{\noindent\textbf{#1}}

\newcommand{\vecx}{\textbf{x}}
\newcommand{\vecy}{\textbf{y}}
\newcommand{\vecp}{\textbf{p}}
\newcommand{\vecn}{\textbf{n}}
\newcommand{\vect}{\textbf{t}}
\newcommand{\vecl}{\textbf{l}}
\newcommand{\vecs}{\textbf{s}}
\newcommand{\vecw}{\textbf{w}}
\newcommand{\vecu}{\textbf{u}}
\newcommand{\vecv}{\textbf{v}}
\newcommand{\vecz}{\textbf{z}}
\newcommand{\vecf}{\textbf{f}}
\newcommand{\vecg}{\textbf{g}}
\newcommand{\vech}{\textbf{h}}
\newcommand{\veca}{\textbf{a}}

\newcommand{\sigRGB}{\vecf}
\newcommand{\sigTonal}{\vecg}
\newcommand{\sigColorTonal}{\vecp}
\newcommand{\sigTonalTrans}{\hat{\vecg}}
\newcommand{\sigHalftone}{\vech}
\newcommand{\sigErrDiff}{\tilde{\vecg}}
\newcommand{\matAssign}{m}

\newcommand{\R}[1]{\mathbb{R}^{#1}}

\newcommand{\norm}[2]{\left\|#1\right\|_{#2}}

\newcommand{\shapeS}{\mathcal{S}}
\newcommand{\surfaceS}{\partial\shapeS}

\newcommand{\labelset}{\mathcal{L}}
\newcommand{\numlabels}{L}

\newcommand{\materials}{\mathcal{M}}
\newcommand{\numMaterials}{M}

\newcommand{\boundingVoxels}{\mathcal{B}}
\newcommand{\voxelSet}{\mathcal{V}}
\newcommand{\surfaceVoxels}{\mathcal{V}_{\partial}}

\newcommand{\transfer}{\textbf{T}}

\newcommand{\layerThick}{\tau}
\newcommand{\voxelLayer}[1]{\voxelSet_{#1}}

\newcommand{\edgeSet}{\mathcal{E}}

\newcommand{\window}{\mathcal{W}}
\newcommand{\nbrs}{\mathcal{N}}

\newcommand{\maxdist}{d_{\max}}

\newcommand{\otherwise}{\mbox{otherwise}}
\newcommand{\suchthat}{\mbox{such that}}

\newcommand{\floor}[1]{\left\lfloor #1 \right\rfloor}

\newcommand{\nearint}[1]{\mbox{round}\left(#1\right)}

\newcommand{\atan}{\mbox{atan2}}

\newcommand{\mm}{\ \mbox{mm}}
\newcommand{\cm}{\ \mbox{cm}}
\newcommand{\m}{\ \mbox{m}}

\newcommand{\ColorSpace}{\ensuremath{\bm{\mathcal{C}}} }
\newcommand{\TonalValueSpace}{\ensuremath{\bm{\mathcal{T}}} }

\newcommand{\etal}{\mbox{\emph{et al.\ }}}

\begin{abstract}
Accurate color reproduction is important in many applications of $3$D printing, from design prototypes to $3$D color copies or portraits. Although full color is available via other technologies, multi-jet printers have greater potential for graphical $3$D printing, in terms of reproducing complex appearance properties. 
However, to date these printers cannot produce full color, and doing so poses substantial technical challenges, from the shear amount of data to the translucency of the available color materials. In this paper, we propose an error diffusion halftoning approach to achieve full color with multi-jet printers, which operates on multiple isosurfaces or layers within the object. We propose a novel traversal algorithm for voxel surfaces, which allows the transfer of existing error diffusion algorithms from $2$D printing. The resulting prints faithfully reproduce colors, color gradients and fine-scale details.
\end{abstract}

\section{Introduction}
\label{sec:intro}

\begin{figure*}[!t]
\centering
	\includegraphics[height=0.25\textwidth]{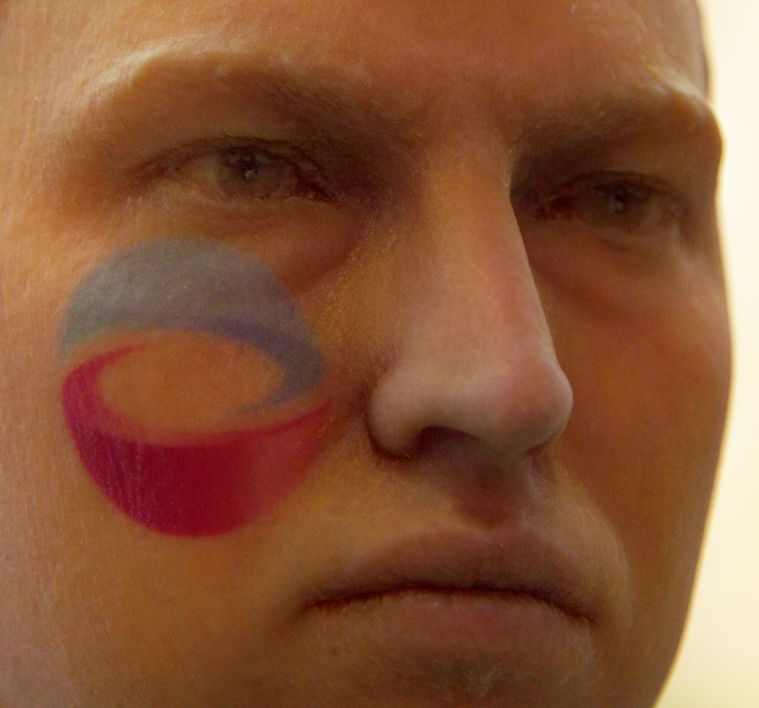}\hfill
	\includegraphics[height=0.25\textwidth]{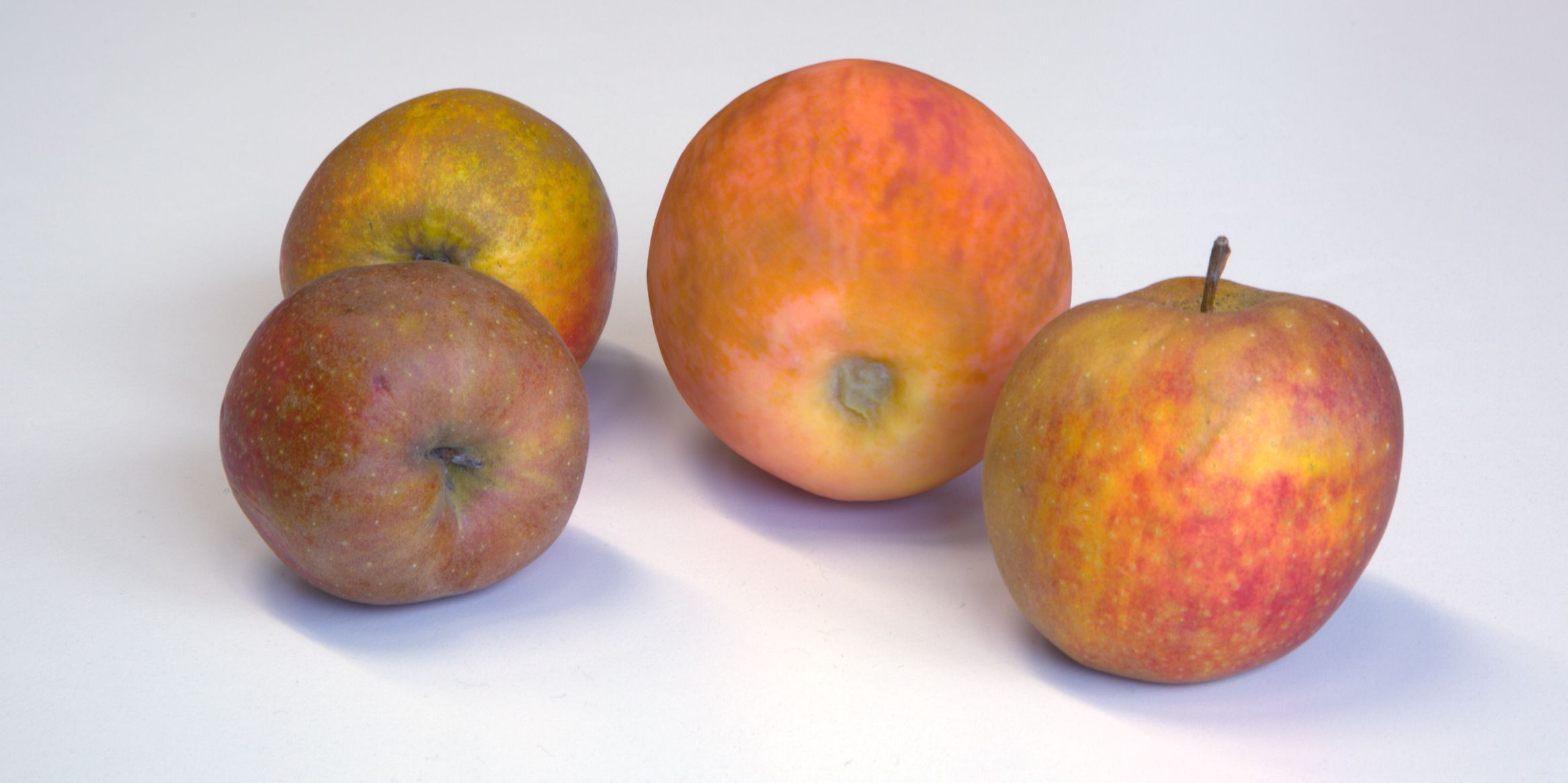}\hfill
	\includegraphics[height=0.25\textwidth]{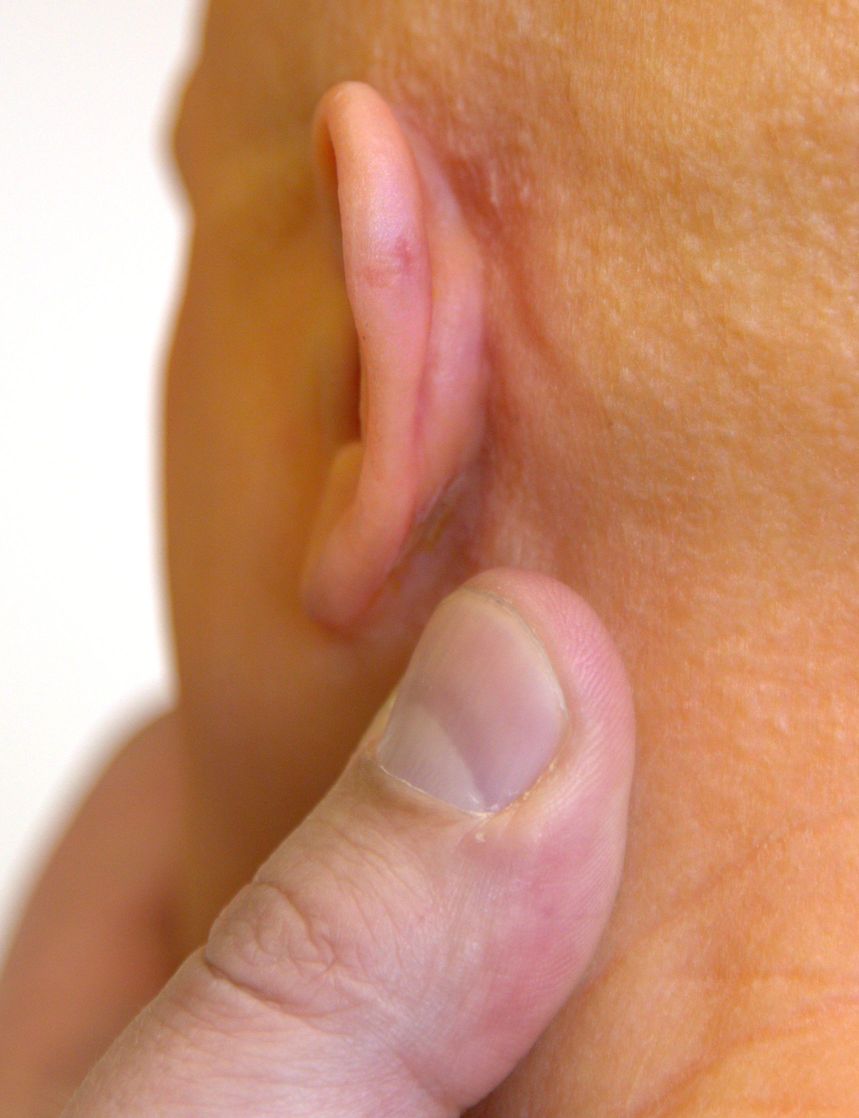}
	\caption{Color 3D prints computed with our software and printed with a multi-material printer show tremendous details and realism. Three of the apples and the thumb are not printed.}
	\label{fig:teaser}
\end{figure*}

In this paper, we present algorithms for producing 3D color prints, which are highly accurate and detailed. From a technical standpoint, we consider the problem of halftoning a signal defined on a surface for the purposes of accurate color reproduction in $3$D printing when using translucent materials. More specifically, we present algorithms for error-diffusion on voxel representations of surfaces, which are compatible with translucent printing materials.

Accurate color reproduction is important in many applications of $3$D printing, especially for design-prototypes or $3$D color copies, where a texture-mapped $3$D scan of an object is to be reproduced in a color-consistent way; this includes $3$D portraits.

While $3$D color printing has existed for some time using powder-binder~\cite{Projet860} and layer-laminated~\cite{Iris} systems, multi-jet $3$D printers with sufficient materials for color printing have only recently come on the market~\cite{Connex3}. In comparison to powder-based systems, multi-jet printers produce smoother surfaces at higher resolutions, and allow more control over the internal structure of the print. In the longer term, such printers have much greater potential to reproduce complex appearance properties. In fact, multi-jet printers with fewer materials have been used to approximate desired single-tone subsurface scattering properties~\cite{Hasan2010,Dong2010}, and opaque objects inside transparent shells~\cite{Vidimce_openfab_2013}. Unfortunately, software allowing full color printing with multi-jet printers is not yet available; the factory software allows colors to be selected from a palette of $\approx 50$ and manually assigned to shells.

Without accurate color reproduction, more complex appearance characteristics are limited in their applicability and appeal. Therefore, accurate color reproduction is a crucial step to fully exploiting the appearance capabilities of these printers. We provide precisely this critical building block for graphical $3$D printing with this paper.

However, there are substantial technical challenges in color reproduction with multi-jet printers. First, is the amount of data that must be processed. For accurate color, the material assigned to each voxel must be controlled. Modern $3$D printers have a resolution of more than $18$ million voxels per cubic centimeter. Thus, prints such as those in Figure \ref{fig:teaser} require tens of billions of voxels (see Table \ref{tab:performance} for details). Both resolution and build volume of $3$D printers are expected to increase in the coming years.

Second, for multi-jet printers currently on the market, the colored materials are highly translucent, which means that the organization of the materials a significant distance beneath the surface, up to a few millimeters, can greatly affect the perceived color as discussed in Section \ref{sec:prelim}. This translucency leads to both blurring of fine-scale details in textures, and to artifacts caused by severe dot-gain if care is not taken in the material arrangement. It is therefore highly important to control material placement several layers of voxels beneath the surface, considering the transmission and scattering properties of the materials, greatly complicating the computational aspects of halftoning algorithms.

While we expect less translucent materials to come on the market in the future, not only will our proposed technique also work for these materials, it will actually perform better--producing equal or larger color gamuts with less computational effort. Further, employing translucent materials may have its own advantages, when we consider materials with similar translucency as observed in skin.

Third, while powder-binder and layer-laminated systems, like $2$D color printing, may combine up to $3$ inks at a single surface location, multi-jet printing allows exactly one material per voxel.

In this paper, we leverage the knowledge of decades of research in color imaging, color management and $2$D color printing, to maximize the quality and exploit the full capabilities of high-resolution multi-material 3D printers--and push their limits towards realism. To this end, we propose a geometry-adaptive error diffusion halftoning algorithm, which includes the following technical contributions:
\begin{itemize}
	\item A traversal algorithm for voxel representations of surfaces, which maps 2D anisotropic error diffusion filters onto a surface in a consistently oriented way (Section \ref{sec:traversal}) and requires only local information to do so.
	\item A layered halftoning algorithm, which combines the traversal algorithm with an arbitrary 2D error diffusion algorithm, and can adapt to the translucency of the materials or increase the color gamut by varying the number of layers (Section \ref{sec:layered}).
\end{itemize}
We do not attempt to approximate the full BSSRDF of an object, but rather present halftoning algorithms capable of reproducing an object's albedo texture even when the printing materials are highly translucent, under non-pathological illuminaton conditions, such as a light source in contact with (as in Figure \ref{fig:material_translucency}) or inside the printed object. Note that a high-quality halftone, such as those produced by our algorithms, is necessary, but not sufficient to accurately reproduce an object's color. Additionally, one needs colorimetrically or spectrally accurate measurements of the object's color, which are typically not provided by common 3D scanners, and an accurate optical printer model. In Section \ref{sec:ColorManagement} we describe a fully empirical model for our printer. A more accurate model is left for future work.

\section{Related Work}
\label{sec:related}

\qheading{Existing 3D color printing technologies.} Powder-binder~\cite{Projet860} and layer-laminated~\cite{Iris} systems provide full color solutions. In both cases, CMYK inks are applied to a white base material. Although these technologies are effective for color printing, they do not provide the resolution or smooth surfaces of multi-jet systems and are limited to nearly opaque materials. They therefore have less potential for future combinations of color with complex appearance properties.

\qheading{Appearance reproduction and multi-material fabrication.} Spec2Fab~\cite{Chen_siggraph2013} is a general reducer-tuner framework for specification-driven digital fabrication, which allows textures to be replicated on a 3D print. They use an error diffusion optimization of material layerings, effectively a contone, with a uniform error filter (error is pushed equally to all neighbors). Although an important first step in texture-mapping for multi-material printers, it does not allow for anisotropic error diffusion filters, and the iterative optimization prohibits a streaming architecture. OpenFab~\cite{Vidimce_openfab_2013} is a programmable fabrication pipeline for $3$D printers, which uses in-slice $2$D error diffusion dithering~\cite{floyd-steinberg}. The authors observe that by dithering in $3$D they could avoid streaks. Our approach treats the color signal where it is defined--on the surface--by mapping $2$D filters into the tangent space of the surface. As discussed below, this allows us to colorimetrically characterize the 3D printer in a geometry-independent way.

Multi-material $3$D Printing has been used to reproduce specified subsurface scattering properties~\cite{Hasan2010,Dong2010}. Fabrication of directional BRDFs for planar or near-planar surfaces has been done using multi-material printing~\cite{Lan2013} and photolithography~\cite{Levin_siggraph2013}.

Given the work that has already been done using multi-material $3$D printing to reproduce complex appearance properties, the introduction of full color opens up tremendous possibilities for future appearance fabrication.

\qheading{Tone reproduction in FDM prints.} Some recent work has focused on the challenging task of improving tone reproduction in fused deposition modeling (FDM) printing. Hergel and Lefebvre~\cite{hergel2014} optimize seam placement in multi-filament FDM prints to hide or reduce artifacts from changing filaments. Reiner et al.~\cite{reiner2014} perform a type of halftoning for FDM printers while maintain long filament paths. Switching filament heads not only creates artifacts, but also increases print time. Both of these methods are specific to FDM printers.

\qheading{3D Halftoning.} $3$D Halftoning has been applied to material composition using $3$D error diffusion filters~\cite{lou_stucki_1998,doubrovski2014} and $3$D dispersed-dot dithering~\cite{ChoSachsPatrikalakisTroxel2003}. For color and appearance reproduction, 3D error diffusion is not appropriate because material assignments closer to the surface will have a greater influence on the appearance of the object than material assignments deeper within the object. Thus, a 3D error diffusion filter would have to adjust its orientation during traversal to account for this and maintain a consistent orientation with respect to the surface. An isotropic filter would produce similar artifacts as are observed with isotropic filters in 2D error diffusion. In contrast, our approach of halftoning on multiple offset surfaces within the object, in addition to the surface itself, results in a halftone that inherently accounts for the geometry of the surface. The relative influence of voxels at different depths from the surface is calibrated in an offline process and built into an International Color Consortium (ICC) profile. Such an offline color calibration process would be very challenging for a 3D filter, because it would require calibrating every possible surface orientation.

\qheading{2D Halftoning.} Generations of researchers in the field of 2D halftoning focused on finding methodologies to optimally arrange printed dots for maximizing print quality (by preserving tone and structure and shifting quantization errors to the highest spatial frequencies possible - see Section \ref{sec:PerceptionPrinting}) subject to technical limitations of printing systems (e.g. the ability to accurately deposit isolated dots)~\cite{LauArce2001}. 

One category of algorithms, called \emph{point processes}, allow a very fast computation of the halftone screens by thresholding pixels using a precomputed threshold mask that is tiled over the 2D image. The traditional clustered-dot order dithering to create amplitude modulated screens and dispersed-dot ordered dithering~\cite{Bayer1973} fall into this category. The latter technique was adapted to 3D printing~\cite{ChoSachsPatrikalakisTroxel2003} accounting particularly for dot placement limitations of binder-jetting systems. One drawback of dispersed-dot ordered dithering is that frequency components given by the screen period are visible resulting in cross-hatch pattern artifacts. Avoiding such artifacts in point process techniques, requires large threshold masks -- e.g. blue-noise masks~\cite{MitsaParker1992} or green-noise masks~\cite{LauArceGallagher1999} -- which are heavily distorted if applied on surface manifolds with non-zero Gaussian curvature.

A second category of algorithms, called \emph{neighborhood processes}, considers a local pixel neighborhood for thresholding. Error-diffusion methods (see Appendix) belong to this category and were pioneered by Floyd-Steinberg~\cite{floyd-steinberg}. Since then, the first error diffusion techniques affected by visual artifacts from low-frequency structural patterns) were improved drastically producing visually pleasing halftones without artifacts. Multiple modifications have been proposed including edge enhancement~\cite{EschbachKnox1991}, tone-dependent diffusion filters~\cite{ostromoukhov_errdiff_2001,Li_tded_2004}, threshold modulation~\cite{ZhouFang_varcoeff_threshmod_2003} structure preservation~\cite{Chang_structaware_ed_2009} or memory efficient processing~\cite{Chang_memeffic_ed_2003}. 

A last category employs models of the human visual system to optimize halftone pattern iteratively \cite{AgarAllebach2005,PangQuWongCohen-OrHeng2008}. Such methods can produce the highest halftone quality, but are in general computationally much more expensive than methods belonging to the other categories. Furthermore, these methods iterate over the whole data and require substantial modifications for adapting them into a streaming architecture. Since iterative methods aim to preserve local structure, their quality advantages over other methods diminish with increasing print resolution. 

Due to the low computational effort and high quality of 2D error diffusion achieved with small diffusion filters, we decided to adapt it to 3D color printing. One prior work addresses error diffusion on a surface~\cite{Alexa2015}. This approach operates on meshes and traverses the vertices based on the availability of subsequent moves or neighbors to diffuse error to. While this approach could be applied to any graph structure, including voxels, it is not clear that it can be applied in a streaming architecture. To the best of our knowledge, we are the first to consider error diffusion halftoning in the context of both non-Euclidean domains and highly translucent materials. Moreover, we are the first to propose such a technique demonstrated to be applicable in practice to tens of billions of elements.

\section{Background}
\label{sec:prelim}

\subsection{Light scattering and absorption in printing materials} \label{sec:LightScatteringAbsorption}

\begin{figure}
\includegraphics[width=0.225\textwidth]{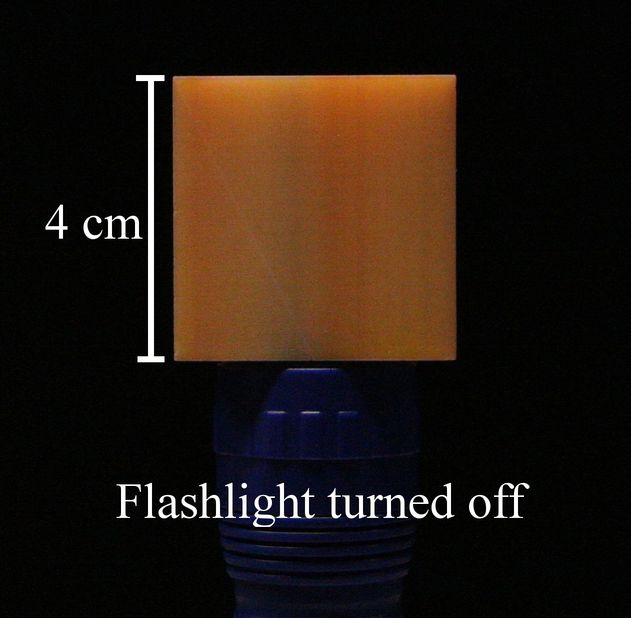}\hfill
\includegraphics[width=0.225\textwidth]{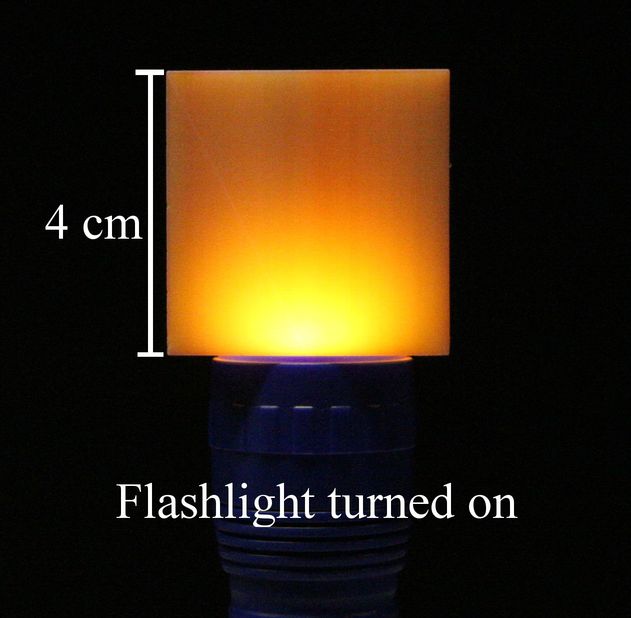}
\caption{Subsurface light scattering in photo-polymer printing materials.}
\label{fig:material_translucency}
\end{figure}

Given an arrangement of multiple non-fluorescent printing materials with similar refractive indexes within a shape $\shapeS$, light propagation within this shape can be described by the steady-state radiative transfer equation~\cite{Chandrasekhar1960} (see supplemental material for details), which shows that the intensity of light traveling through the material is attenuated by absorption (dependent on wavelength, independent of direction) and is redistributed by scattering. Thus, a fraction of light entering the print at one location may be emitted from the surface at a different location due to scattering (see Figure \ref{fig:material_translucency}). If light travels through different materials its spectral power distribution is modulated by each material's absorption coefficients and the path length within this material.

This has multiple implications for arranging translucent (high scattering, low absorption) printing materials colored in cyan (C), magenta (M), yellow (Y) and white (W) for full color 3D printing.

\emph{Highest Reflectance (white point):} To maximize reflectance, not only the surface must be covered with white material, but also the space beneath. A significant fraction of light travels deep into the object due to the low absorption. Each non-white material with higher absorption placed into the light path would absorb light that would otherwise contribute to the albedo. There exists a minimum thickness $d_w$ of a white layer to create almost maximum reflectance, i.e. for each $d > d_w$: $\left\|R_{d_w} - R_{d}\right\|_2 < \Delta R_{\max}$, where $R_{d_w}$ and $R_{d}$ are spectral reflectances of white layers with thicknesses $d_w$ and $d$, and $\Delta R_{\max}$ is the maximally acceptable reflectance difference.

\emph{Lowest Reflectance (black point):} To minimize reflectance, the printing material arrangement must ensure that light in the whole visible wavelength range is maximally absorbed. This can only be achieved by appropriately arranging C, M, Y materials on the surface and beneath. The minimal layer thickness $d_b$ of such an arrangement for creating almost minimal reflectance (deviating only by $\Delta R_{\max}$ from minimum) satisfies $d_b < d_w$ because the color materials' absorption coefficients are larger than that of white (assuming similar scattering properties of all materials). 

Therefore, not all voxels in the shape addressable by the printer need to be considered. We can drastically reduce computation by filling all voxels within the shape by default with white (maximizing reflectance of the white point) and compute the arrangement of remaining materials only within $d_b$ of the surface. Although material placement with a distance $d \in [d_b,d_w]$ from the surface may impact reflectance, it affects only very bright colors which are also reproducible by color halftones in upper layers. While theoretically one could derive $d_b$ based on measurements of the material parameters, Section \ref{sec:layered:construct} shows how selecting a maximum distance $\maxdist \leq d_b$ from the surface allows an empirical trade-off between the number of voxels that must be addressed and the achievable gamut volume.

\begin{figure}
\centering
\fbox{\includegraphics[width=0.23\textwidth]{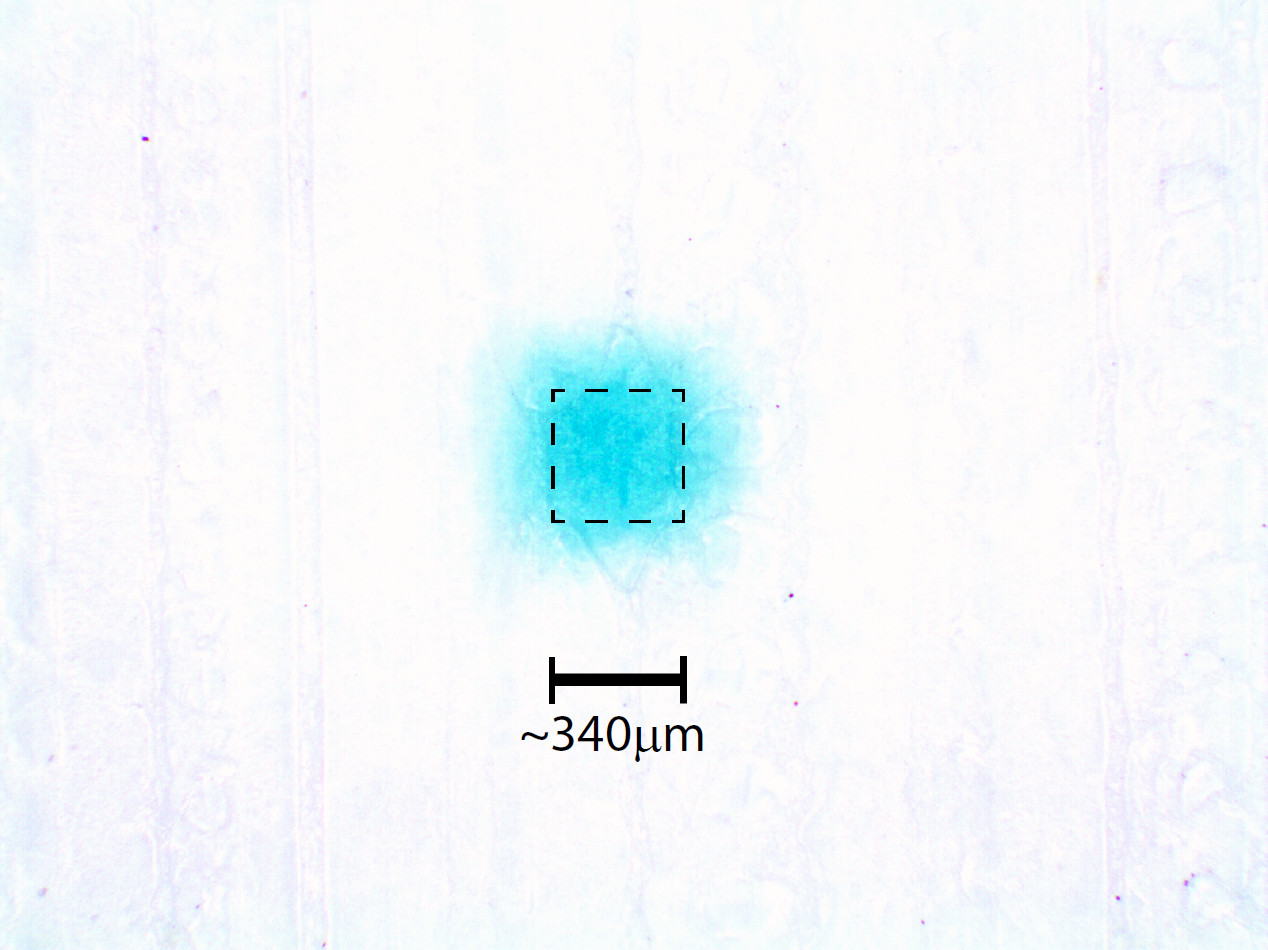}}\hfill
\fbox{\includegraphics[width=0.23\textwidth]{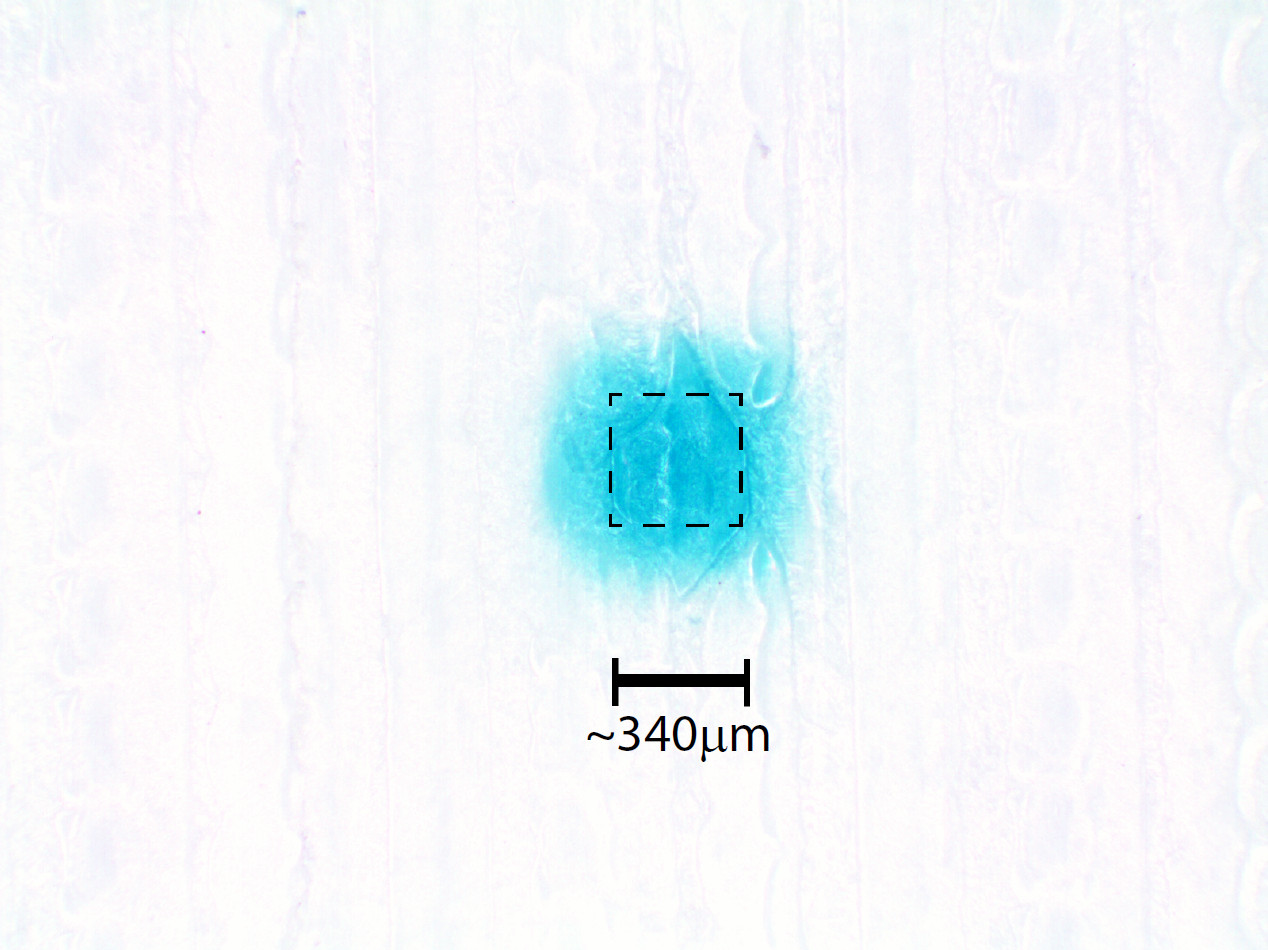}}
\caption{Printing with translucent materials: severe dot gain and increased contrast by printing more slices. Two small cyan printed squares (indicated by a thin dashed line), both $8$ dots or $\sim\! 340\ \mu\mbox{m}$ wide at $600$ DPI. Left: the square is printed $9$ slices deep into surrounding white material. Right: $36$ slices deep. Also visible is the surface topography due to printing roller.}
\label{fig:dot_gain}
\end{figure}

\emph{Optical Dot Gain:} Voxels occupied by a colored material and surrounded by white material appear larger because the spectral power distribution of light traveling through both materials is modulated mostly by absorption of the colored material. This phenomenon is known in 2D printing as \emph{optical dot gain}~\cite{Rogers1997}. The more colored voxels are stacked on top of each other aligned with the surface normal the more laterally scattered light is modulated by the colored material. Thus, the colored area on the surface appears not only darker, but also larger as shown in Figure \ref{fig:dot_gain}. Algorithms that use surface halftones to assign materials to inner voxels must consider this phenomenon to avoid dot gain artifacts particularly visible in bright areas by isolated disturbing large dots. Section \ref{sec:layered:assign} describes how independently halftoning layers allows to decorrelate material assignments in different layers, which greatly reduces dot gain artifacts in translucent materials.

\setlength{\fboxrule}{0.5mm}
\setlength{\fboxsep}{0mm}

\subsection{Perceptual aspects in color printing} \label{sec:PerceptionPrinting}
Each printing system is limited in reproducing reflectances due to the small number of available printing materials. Still, accurate color reproduction is possible by material arrangements creating physical errors for which the human visual system (HVS) is insensitive. Two perceptual aspects are particularly important in printing.

\emph{Contrast perception:} Contrast sensitivity functions (CSFs) describe the HVS's sensitivity to achromatic and chromatic contrasts as a function of spatial frequency, orientation and luminance~\cite{Mullen1985,CampbellKulikowskiLevinson1966,VanNesBouman1967}. The achromatic CSF has bandpass and chromatic CSFs have lowpass characteristics. Each CSF monotonically decreases for frequencies higher than 10 cycles/degree resulting in an apparent blurring of high frequency contrasts. This is used in printing by arranging printing materials in high frequency patterns (halftones) to create various color shades from only a few colored printing materials.

\emph{Color perception:} Retinal processing reduces the information content of a spectral stimulus to only three values--the cone responses~\cite{WyszeckiStiles2000}. Therefore, different reflectance spectra viewed under the same illuminant match visually if corresponding spectral stimuli yield similar cone responses. By agreeing on the viewing illuminant, accurate color printing is possible by material arrangements that reproduce cone responses instead of reflectances. A drawback of this \emph{metameric color reproduction} is a likely mismatch between prints and originals for other illuminants.
CIE colorimetry provides two standardized color spaces for metameric printing (see e.g.~\cite{Fairchild2013}):
CIEXYZ, which is linearly related to cone responses; and CIELAB, an opponent color space derived from CIEXYZ that allows simple access to color attribute predictors for lightness, chroma and hue and is perceptually more uniform. Color control in metameric printing is done by \emph{separation}--relating CIEXYZ or CIELAB values to material arrangements. Since this is generally impossible for all colors due to gamut limitations of the printing system, a preceding \emph{gamut mapping} method must transform all colors into the device gamut aiming to minimize perceptual errors between original and reproduction~\cite{Morovic2008}.

\section{Overview}
\label{sec:overview}

\begin{figure*}[ht]
\includegraphics[width=\textwidth]{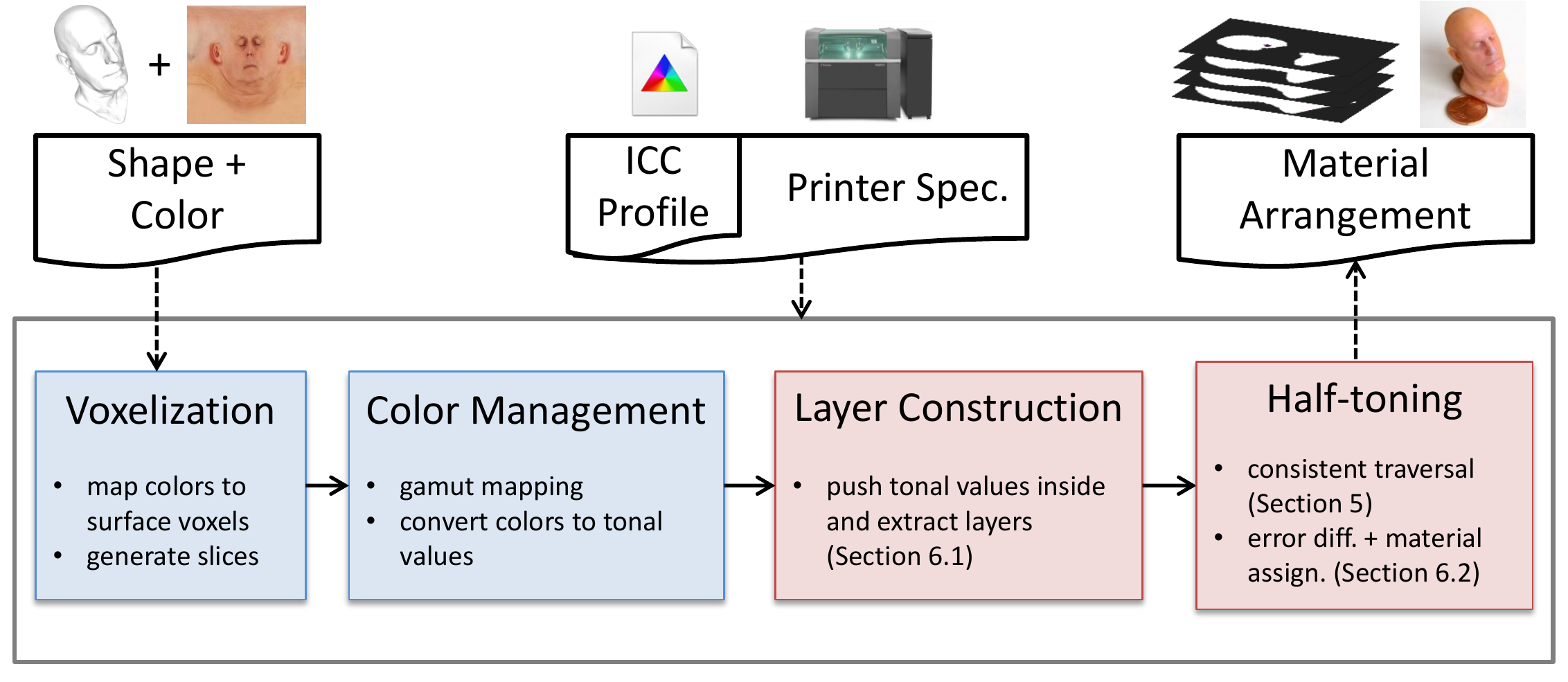}
\caption{Flow-chart of our processing pipeline.}
\label{fig:flowchart}
\end{figure*}

Given the resolution of current 3D printers, and expected increases in both resolution and build space, we need a streaming model for computation, wherein only localized data and computation are needed. 
Figure \ref{fig:flowchart} shows such a streaming framework, which takes as input shape and color information, and produces printer compatible data in which a single material is assigned to each voxel.
The error diffusion and material assignment algorithms described in Sections \ref{sec:traversal} and \ref{sec:layered} are our core contributions and take place in the parts colored reddish. Each aspect of our framework is described briefly below.

\qheading{Shape $+$ Color:} 
The input to our pipeline is a shape $\shapeS\subset\R{3}$ defined by a surface $\surfaceS$ (e.g. tessellated mesh) with attached color information $\sigRGB:\surfaceS\mapsto \ColorSpace$ (texture data or per-vertex colors), where \ColorSpace is specified by an invertible transformation from CIEXYZ defined for an illuminant (e.g. CIED50) and a set of color matching functions (CMFs) (e.g. CIE 1931 CMFs). Examples for \ColorSpace are standard RGB color spaces~\cite{SuesstrunkBuckleySwen1999}. 
By default we interpret RGB data as sRGB.

\qheading{Voxelization:} 
The axis-aligned bounding box of our shape $B(\shapeS)$ is discretized with a fixed, regular grid of voxels $\boundingVoxels$ at the resolution given by the printer specification. We slice the surface and identify interior, $\voxelSet=\shapeS\cap\boundingVoxels$, and surface voxels, $\surfaceVoxels\subset\voxelSet=\surfaceS\cap\boundingVoxels$, where a \emph{slice} refers to a 2D array of voxels with a constant $z$-coordinate. The full slice is denoted by $\boundingVoxels(s)\subset\boundingVoxels$ for slice $s$ with center $z$-coordinate $z_s$, and $\surfaceVoxels(s)=\surfaceVoxels\cap\boundingVoxels(s)$ denotes the surface voxels in slice $s$. Our voxelizer generates slices in chunks of $\approx 100$ slices--$\approx 3\mm$ for our hardware setup--depending on the number of layers and layer thickness as discussed in Section \ref{sec:layered}, until the object is completely sliced. Each chunk proceeds through the pipeline until materials have been assigned, and printer-specific output has been generated. Then the next chunk is generated.

During the voxelization process, we assign colors to the surface voxels; abusing notation slightly, we redefine $\sigRGB:\surfaceVoxels\mapsto \ColorSpace$ as a function on the surface voxels. We sample the texture data using trilinear interpolation (mipmapping) with the level-of-detail computed per surface voxel. This is key to avoid aliasing--while there are no viewpoint-dependent effects as in rendering, the texture map is in general non-uniform over the surface. 

\qheading{Color Management:} In this stage, which includes gamut mapping and color separation, we map color data to printer tonal values using ICC color management \cite{ICC43} (see section \ref{sec:ColorManagement}): $\sigColorTonal:\ColorSpace \mapsto \TonalValueSpace$, where each element of the tonal value space $\TonalValueSpace$ corresponds to the amount of available printing materials required to best reproduce a desired color. By function composition we attach one tonal value vector to each surface voxel: $\sigTonal = \sigColorTonal \circ \sigRGB: \surfaceVoxels \mapsto \TonalValueSpace$. 

\qheading{Layer Construction:}
To account for the translucency of the materials, see Figures \ref{fig:material_translucency} and \ref{fig:dot_gain}, we transfer the tonal values stored in the surface voxels to the interior voxels within a distance $\maxdist$ of the surface. See Figure \ref{fig:layer_illustrate} for an example. These voxels have the greatest influence on the appearance of the printed object. From these voxels, we extract a set of layers as isosurfaces of distance to the surface of the object. The number and thickness of the layers determine $\maxdist$, and these are chosen as described in Section \ref{sec:layered:construct}.

\qheading{Halftoning:} We then treat each layer as a distinct surface with tonal values, and halftone them independently using the surface traversal algorithm described in Section \ref{sec:traversal} and the error diffusion and material assignment algorithm described in Section \ref{sec:layered}.

\qheading{Material Arrangement:}
The output of the halftoning is a voxel-level material arrangement $\matAssign:\voxelSet\mapsto \materials$, where the set of materials $\materials$ is discrete and small--$4$ in our case. We store $\matAssign$ per-slice, and convert it to a printer-specific format to send to the device.

\section{Consistent Surface Traversal}
\label{sec:traversal}

\newcommand{\distToEmpty}[1]{\phi_{#1}}

Classical error diffusion algorithms for 2D images, e.g.~\cite{floyd-steinberg,ostromoukhov_errdiff_2001}, rely on the partial ordering property of subsets of Euclidean domains, to define a traversal order, eg. raster scan or serpentine scan order, in which error is never pushed to pixels that have already been halftoned.

On a surface, which has non-zero curvature and is in general not a topological disc, such a partial ordering property does not exist without introducing a seam or a singularity, a consequence of the Hairy Ball Theorem. Any traversal scheme will eventually encounter an element that has already been halftoned, requiring either a stop and restart, or a change in direction. Excessive stopping and restarting can lead to poor distribution of error, and we design a traversal that avoids this. 

To map anisotropic error diffusion filters on to the surface requires a local coordinate system of the tangent plane at a given point, which is fixed with the surface normal and one additional direction. Mapping a filter to the surface so that it is consistently oriented for adjacent points is equivalent to parameterizing the tangent space of a surface in a consistent way, or finding smooth vector fields on a surface, which is in general an ill-posed problem and finding an optimal solution requires considering the entire surface.

Instead, we propose a voxel-level traversal algorithm that maintains a consistent orientation of the filter, traverses the surface voxels in long runs allowing error to accumulate and diffuse evenly, and does so using only local information. Our traversal avoids situations where voxels only distribute or obtain error during diffusion. 

We maintain a consistent orientation of the filter by traversing the voxels in a consistent direction, and using the traversal direction to fix a coordinate system in the tangent plane. While our traversal provides no theoretical guarantees, it performs well in practice. 

\subsection{Geometry of Voxel Slices}
\label{sec:traversal:geometry}

As a consistent direction, we always travel clockwise or counterclockwise about a line parallel with the positive $z$-axis. For brevity, we describe only the counterclockwise case. Note that we want to traverse counterclockwise, not about a global axis, but rather about a line through the center of the local \emph{connected component} of voxels. Depending on the geometry of $\shapeS$, when it intersects a slice plane $z=z_s$ and is discretized into a voxel representation, the resulting voxels in slice $s$ may belong to multiple connected components. An example of a slice with multiple connected components can be seen in Figure \ref{fig:layer_illustrate}.

\begin{figure}
\centering
\includegraphics[width=0.475\textwidth]{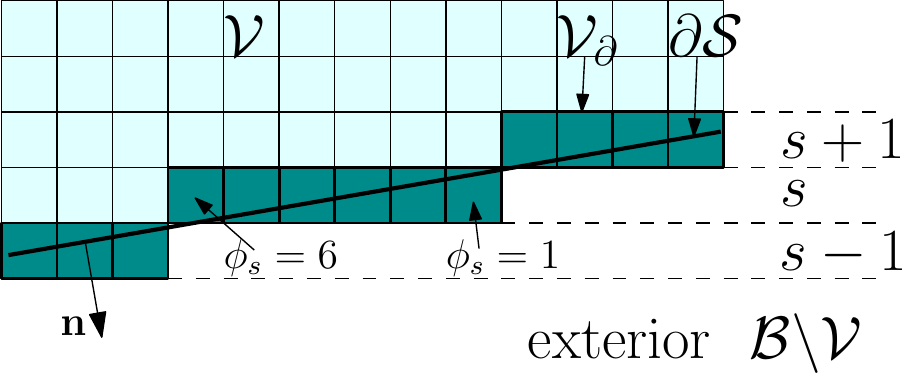}
\caption{Geometry of surface voxels in consecutive slices for a down-facing surface.}
\label{fig:slice_geom}
\end{figure}

If the surface normal is close to vertical (almost parallel with the positive or negative $z$-axis), then $\surfaceVoxels(s)$ will be multiple voxels wide. These surface voxels $\surfaceVoxels(s)$ may enclose interior voxels in slice $s$, and will be themselves enclosed by exterior or empty voxels in slice $s$. Figure \ref{fig:slice_geom} shows a 2D cross-section of this. 

If we consider a part of the shape, which forms a connected component in consecutive slices, and where the surface faces almost in the negative $z$ direction, but slopes or curves slightly upward, we have the following situation, which is shown in 2D in Figure \ref{fig:slice_geom}. When projected into the $xy$-plane, the surface voxels of $\surfaceVoxels(s)$ will be located inside the surface voxels of $\surfaceVoxels(s+1)$.
If we consider two surface voxels within the same connected component in slice $s$, the innermost should be traversed first, since they are next to $\surfaceVoxels(s-1)$ and have already received error diffused from slice $s-1$. The outermost should be traversed last, which are next to $\surfaceVoxels(s+1)$.
The situation is reversed for upward facing surfaces.

\subsection{Traversal Algorithm}
\label{sec:traversal:algorithm}

\begin{figure*}[!t]
\centering
\includegraphics[width=0.975\textwidth]{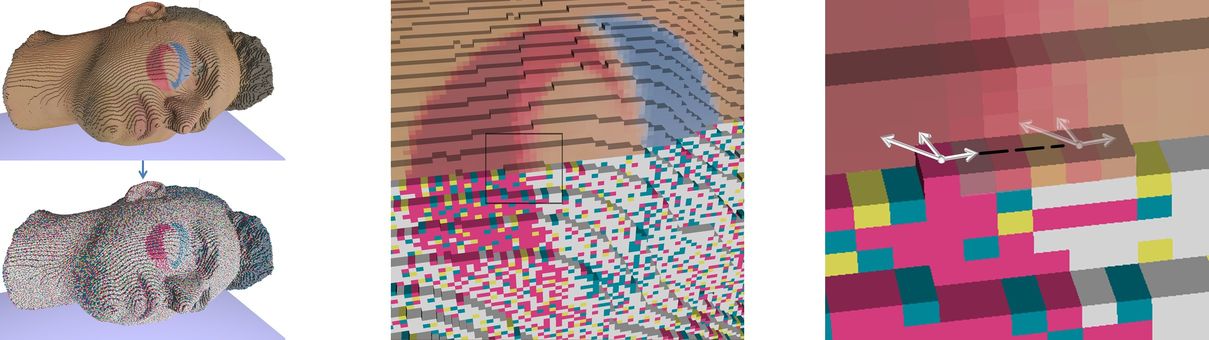}
\caption{Traversal of a surface maintaining consistent orientation of an error diffusion filter. Left: a rendering of a $1\cm$ voxelization of a head model before and after halftoning. Middle: a close-up of the halftoning process up to a given slice. Right: a cut-out from the middle of how a 3-component filter moves as each voxel is quantized. Note: the halftone is shown as opaque and diffuse for illustration purposes, and does not reflect the printed appearance.}
\label{fig:traversal}
\end{figure*}

We construct an undirected graph for traversal as follows: for each surface voxel $\vecv\in\surfaceVoxels$, we connect it with an edge to each other surface voxel within the $3\times3\times3$ voxel window $\window(\vecv)$ centered on $\vecv$. We denote this set of one-ring neighbors $\nbrs(\vecv)\subset\surfaceVoxels$. We traverse the voxel representation of the surface slice by slice, analogous to halftoning an image row by row, by selecting the next voxel from the set of neighbors that are within the same slice, $\nbrs_{s}(\vecv)\subseteq\nbrs(\vecv)$. Error is always diffused upwards to the next slice, and to other surface voxels within the slice, which have not yet been assigned materials, as shown for a very small example ($1\cm$) in Figure \ref{fig:traversal}. 

One might also consider standard graph traversals, such as breadth-first or depth-first, also constrained to neighbors within the same slice, but these do not maintain a consistent direction of travel and therefore only permit symmetric error diffusion filters (e.g. uniform, Gaussian, etc.).

As described in Section \ref{sec:traversal:geometry}, the traversal winds up the surface on a slice-by-slice basis counterclockwise about a line through the center of each connected component. 
However, finding the center of the current connected component would require a search over all voxels in the slice to determine the connected components, find their centers, and to which connected component each surface voxel belongs. Such a search would prove too expensive--essentially involving a pre-traversal.

We can determine whether a potential next voxel in proper traversal order winds about the connected component in the correct direction using only two pieces of local information. The first is the surface normal. The second is the in-slice distance to the exterior of $\shapeS$, or distance-to-empty, denoted by $\distToEmpty{s}(\vecv)$ for voxel $\vecv\in\surfaceVoxels(s)$
\begin{equation}
\distToEmpty{s}(\vecv) = \min_{\vecu\in\boundingVoxels(s)\backslash\voxelSet(s)} \norm{\vecv - \vecu}{1},
\end{equation}
which can be pre-computed for each slice--see Figure \ref{fig:slice_geom} for an example. 
We use the $L_1$-norm for simplicity and efficiency, but another norm would also work. 
The gradient of $\distToEmpty{s}(\vecv)$ always points to the interior of the connected component to which $\vecv$ belongs. This is important for situations where the current slice may have multiple disconnected components, as in Figure \ref{fig:layer_illustrate}. 

This allows us to efficiently test if a potential next voxel $\vecu\in\nbrs_{s}(\vecv)$ continues in the same direction around the connected component by computing $(\vecu - \vecv)\times \nabla\distToEmpty{s}(\vecv)$. For example, if we are traversing counterclockwise, we can rule out all neighbors of $\vecv$ for which this cross-product has a negative $z$-component. We denote the subset of $\nbrs_{s}(\vecv)$ satisfying this direction criterion $\nbrs_{\times}(\vecv)$.

Referring to Figure \ref{fig:slice_geom} and recalling the observations of Section \ref{sec:traversal:geometry} we get the following additional traversal criterion. If the surface faces downward, we wish to stay as \emph{inside}--away from exterior voxels--as possible. If the surface faces upward, we wish to stay as \emph{outside} as possible. Hence we choose the neighbor 
\begin{equation}
\vecw = \arg\max_{\vecu\in\nbrs_{\times}(\vecv)} \distToEmpty{s}(\vecu)
\end{equation}
for down-facing surfaces, and
\begin{equation}
\vecw = \arg\min_{\vecu\in\nbrs_{\times}(\vecv)} \distToEmpty{s}(\vecu)
\end{equation}
otherwise.

\begin{figure}
\centering
\parbox{0.15\textwidth}{
\centering
\includegraphics[height=3cm]{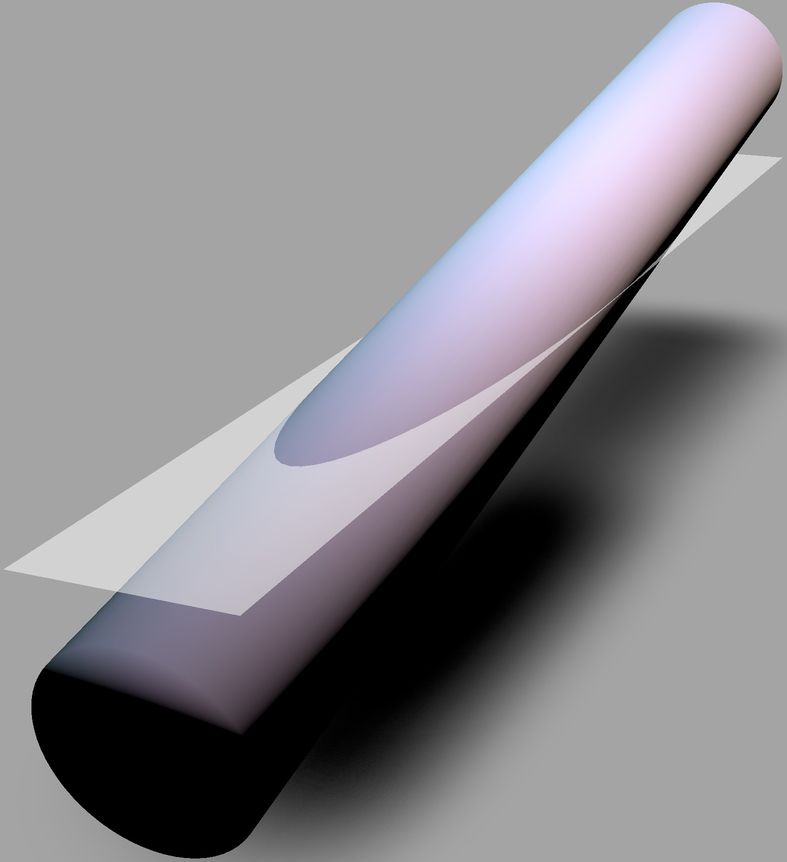}\\
(a)
}
\hfill
\parbox{0.1\textwidth}{
\centering
\includegraphics[height=3cm,width=2cm]{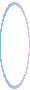}\\
(b)
}
\hfill
\parbox{0.15\textwidth}{
\centering
\includegraphics[height=3cm,width=2cm]{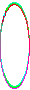}\hfill
\includegraphics[height=3cm,width=0.5cm]{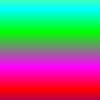}\\
(c)
}
\caption{A visualization of traversal order for a slice containing both upward and downward facing parts: (a) slice plane intersecting the object; (b) tonal values at the surface in that slice; (c) color-coded traversal order with color-coding legend (bottom=first, top=last).}
\label{fig:traversal_order}
\end{figure}

Because the surface orientation is determined locally using the surface normal, the traversal adjusts as it traverses a slice with both up- and down-facing segments. Figure \ref{fig:traversal_order} shows the traversal order for such a slice.

\subsection{Boundary Cases}
\label{sec:traversal:cases}

\qheading{Start point selection.}
We also use the distance-to-empty, in combination with one other local quantity, to select the start point in each slice. First, potential starting voxels (voxels in the slice, which have not yet been traversed) are filtered based on their \emph{error diffusion count}, or number of times they have had error diffused to them from previous slices. We select the subset of voxels with the largest error diffusion count, and subsequently select from this subset using the distance-to-empty. If the $z$-component of the surface normal is negative, the start point is the point, which maximizes the distance-to-empty. If the $z$-component of the surface normal is positive, the start point is the point, which minimizes the distance-to-empty (i.e. one of outer-most points).

Once the start point and traversal direction are selected, the direction is tested to ensure it is not a dead end. If no voxel is available in that direction, the direction is reversed. This reduces the occurrence of single voxels with few neighbors to push error.

\qheading{Birth and death of components.} 
During the slicing process, new parts of the shape are encountered, creating new connected components of voxels in a given slice. When a new component is encountered in a given slice, the surface voxels will form a disc in that slice, meaning it can be traversed like an image. We handle these cases specifically as follows. Rather than spiral out, as with other downward facing surfaces, we select a global axis, $x$ or $y$, and perform a 2D serpentine scan in this direction. This avoids situations where a long strip of voxels mostly has error pushed away from it, creating a start-up artifact noticeable for low tonal values. We also use a 2D serpentine traversal for components, which are ending in the current slice, rather than spiraling in. 

We detect the birth of a component when all possible starting points have an error diffusion count of $0$, and the death of a component when all possible starting points have no neighbors in slice $s+1$.

\qheading{Serpentine Surface Scan.} 
When we can no longer traverse in the same direction locally (all $\vecu\in\nbrs_{\times}(\vecv)$ have been traversed), we reverse direction, as in the serpentine traversal scheme from 2D error diffusion, which is shown to provide significant improvement over raster-scan traversal in images \cite{Ulichney1987,Chang_memeffic_ed_2003}. Further, when selecting a new start point, we set the traversal direction opposite the traversal direction, in which error was last diffused to the new starting voxel.

\subsection{Mapping the filter}
\label{sec:traversal:mapping}

Once we have chosen the next voxel to traverse, we can establish a local coordinate system at the current surface voxel, with which we can align the error diffusion filter and the neighbors of the current voxel. As in image error diffusion, one non-zero filter element is aligned with the next voxel to be traversed. This gives us a direction, which together with the surface normal, gives us the necessary coordinate system for the tangent plane at each surface voxel. Depending on whether the direction is clockwise or counterclockwise, we use either a left-handed or right-handed coordinate system. Figure \ref{fig:traversal} shows how an error diffusion filter with $3$ non-zero elements is mapped to the neighbors in the tangent space of a voxel as it traverses the surface voxels.

This is closely related to decal mapping methods such as discrete exponential maps~\cite{schmidt_dexp_2006}, and to parallel transport on manifolds, and for large decals exponential maps provide a more general solution. However, the filters used for error diffusion are typically small enough that the neighbors can simply be projected into the tangent plane while preserving distances.

We let the error diffusion filter live in the local tangent coordinate system, orthogonally project the neighbors onto the tangent space, and find the best alignment of neighbors and non-zero filter elements. We align neighbors with filter elements using symmetric closest points to avoid a single neighbor being assigned to multiple filter elements. That is, both the filter element and the neighbor must be closest to each other.

\subsection{Analysis}
\label{sec:traversal:analysis}

The distance to empty can be computed in time linear in the number of voxels in each slice using a distance transform~\cite{Felzenzwalb2004}. While this means over the whole print job, every voxel in $\boundingVoxels$ must be visited, the constant hidden in the $O(\cdot)$ notation is very small and the distance transform is highly efficient. Selection of the starting point requires a search over all unprocessed surface voxels, but for non-pathological cases this will only be necessary a small number of times per slice. Each traversal operation requires a constant number of operations, as the number of neighbors is constant. Normals can be computed in $O(|\surfaceVoxels(s)|)$ per slice using a signed distance field as an implicit surface definition, as described in Section \ref{sec:layered:construct}. Thus, the total time complexity to fully process $\shapeS$ is linear in the total number of voxels $O(|\boundingVoxels|)$ with a small constant, which is very well suited to a streaming architecture.

\begin{figure*}[t]
\centering
\parbox[b]{0.225\textwidth}
{\centering
\includegraphics[width=0.225\textwidth]{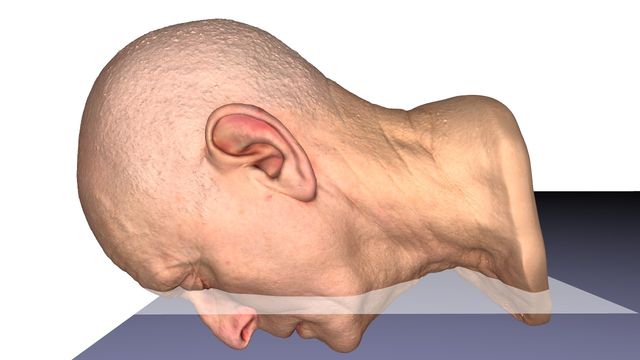}\\
(a)
}\hfill
\parbox[b]{0.225\textwidth}
{\centering
\includegraphics[width=0.225\textwidth]{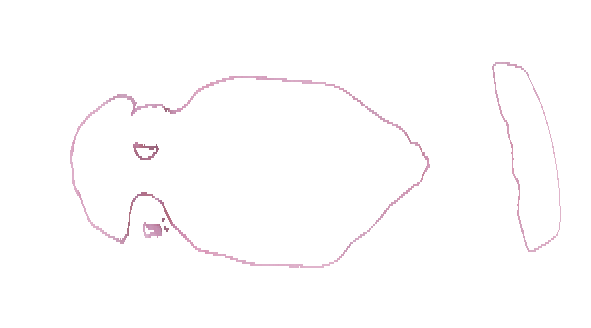}\\
(b)
}\hfill
\parbox[b]{0.225\textwidth}
{\centering
\includegraphics[width=0.225\textwidth]{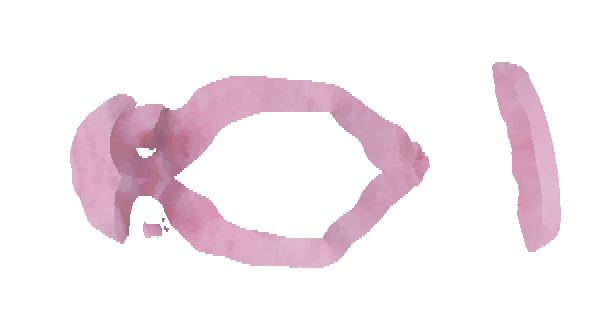}\\
(c)
}\hfill
\parbox[b]{0.225\textwidth}
{\centering
\includegraphics[width=0.225\textwidth]{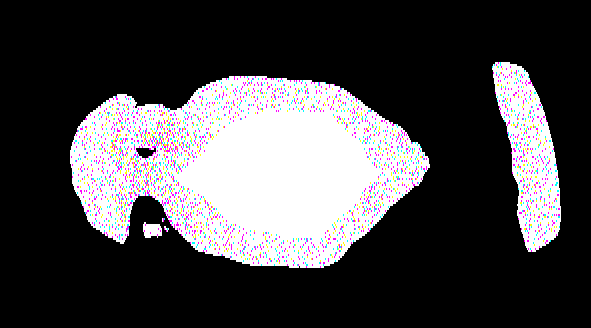}\\
(d)
}
\caption{Layered halftoning process: (a) slice plane $z=z_s$ intersecting the object; (b) the sRGB values in $\surfaceVoxels(s)$ converted to tonal values $\sigTonal$ via the map $\sigColorTonal$; (c) transferred tonal values $\sigTonalTrans$ inside the object; and (d) material assignments $\matAssign$ (black indicates no material).}
\label{fig:layer_illustrate}
\end{figure*}

\section{Layered Halftoning}
\label{sec:layered}

Applying a halftone only to the surface voxels will result in a very low contrast print, due to the translucency of the materials. We can see this from Figure \ref{fig:dot_gain}, which shows how increasing the depth of color material assignment increases the contrast. In Figure \ref{fig:layer_gamut}, we see how increasing the depth of color assignment increases the color gamut. For this reason, we introduce \emph{layered halftoning}, in which color data is transferred from the surface to additional isosurfaces, or layers, within the object, which are then independently halftoned using the traversal scheme presented in Section \ref{sec:traversal} and an adaptive filter~\cite{ostromoukhov_errdiff_2001} and threshold error diffusion algorithm~\cite{ZhouFang_varcoeff_threshmod_2003}.

\subsection{Layer Construction}
\label{sec:layered:construct}

Layered halftoning begins by transferring the surface tonal values $\sigTonal$ from $\surfaceVoxels$ to a subset of $\voxelSet$ within a distance $\maxdist$ of the surface. All interior voxels more distant from the surface are assigned a white material to maximize reflectance as explained in Section \ref{sec:prelim}. This requires both many distance computations and a transfer operator $\transfer : \surfaceVoxels\times\TonalValueSpace \mapsto \voxelSet\times\TonalValueSpace$, which maps functions on $\surfaceVoxels$ to functions on $\voxelSet$.

To compute the distance field $d: \boundingVoxels\mapsto\R{+}$ of voxels to the surface, we leverage the fact that we are only interested in voxels within a given distance $\maxdist$. This allows us to construct a $3$D mask containing these distances at the printer resolution. The distance field $d(\vecv)$ is initialized with a large value $d_{\mbox{\small null}} > \maxdist$. Distance computation then proceeds by moving this mask over the surface voxels $\surfaceVoxels$, and writing the mask value to the voxel at the corresponding offset, if that value is less than the value already there. While this approach is efficient only for relatively small distances, it has the advantage of being embarrassingly parallel and requiring only a less-than operation at each voxel.

As a transfer operator, we simply take the tonal value $\sigTonal(\vecu)$ of the nearest surface voxel $\vecu$, allowing the transfer operator to piggy-back on the distance computation at minimal extra cost, i.e. interior tonal values are assigned as 
\begin{equation}
\sigTonalTrans(\vecv) = \sigTonal(\vecu)\ \mbox{such that}\ \vecu = \arg\min_{\vecs\in\surfaceVoxels} \norm{\vecv - \vecs}{2}
\end{equation}
for $\vecv\in\voxelSet$. Aside from being easy to compute, this has the benefit of preserving high-frequencies in $\sigTonal$. Figure \ref{fig:layer_illustrate} shows the process of assigning sRGB color values to surface voxels, converting them to tonal values, transferring the tonal values to interior voxels and applying error diffusion to get the final material assignment. Note that because tonal values are pushed to the interior in 3D, interior tonal values may include surface tonal values from different slices.

Although there are inevitably discontinuities in $\sigTonalTrans$ at the cut loci of the distance field $d$ (voxels $\vecv$ where multiple surface voxels are equidistant), we found this not to be a problem as the tonal values mostly vary smoothly without significant correlation between high frequencies in $\sigTonal$ and high curvature of the surface. 
In this case we simply assign tonals in a first-come fist-served manner. Similarly, although areas with varying curvature result in slight tone shifts from $\sigTonal$ to $\sigTonalTrans$, in particular for areas of high positive mean curvature, we did not find this to be a problem, even for small prints.

\begin{figure*}
\centering
\parbox{0.25\textwidth}{
\centering
\includegraphics[width=0.25\textwidth]{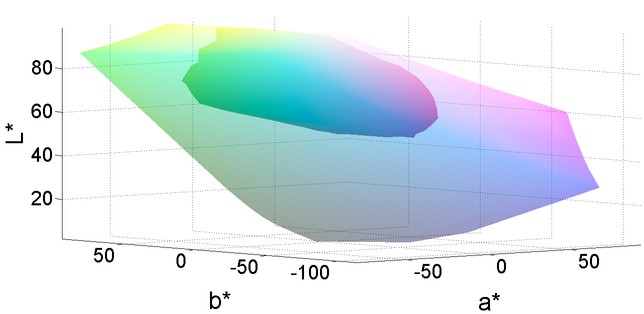}\\0.269
}\hfill
\parbox{0.25\textwidth}{
\centering
\includegraphics[width=0.25\textwidth]{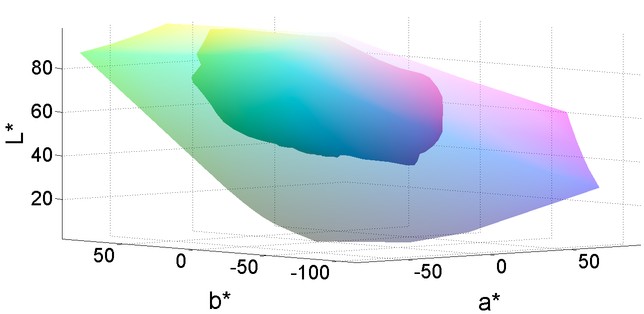}\\0.322
}\hfill
\parbox{0.25\textwidth}{
\centering
\includegraphics[width=0.25\textwidth]{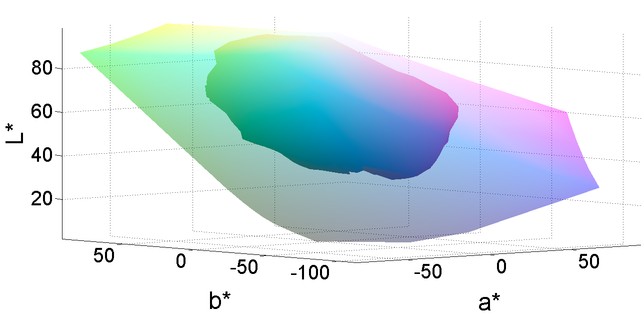}\\0.372
}\hfill
\parbox{0.25\textwidth}{
\centering
\includegraphics[width=0.25\textwidth]{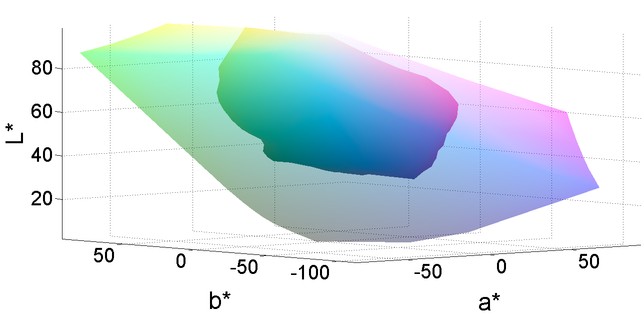}\\0.409
}
\caption{Increasing number of layers results in increasing gamut volume. From left to right: 3, 6, 12 and 18 layers. The gamut volume is indicated below each image as a fraction of the volume of sRGB gamut shown transparent.}
\label{fig:layer_gamut}
\end{figure*}

We extract layers of voxels with tonal values from $\voxelSet$ by defining isosurfaces of $d$ with which to segment $\voxelSet$. As isosurfaces we choose $d_{\ell} = \ell \layerThick$, where $\layerThick$ is the layer thickness, which we choose to be the voxel size along the dimension of minimum resolution. We use $\ell=0,1,\ldots,L-1$, where $L$ is an integer constant, which defines $\maxdist = L\tau$. Ideally $\maxdist = d_b$ (see Section \ref{sec:LightScatteringAbsorption}) to minimize reflectance for maximizing contrast. However, we make a tradeoff between computational effort and gamut volume and set $L = 12$ for all prints shown in this paper. Figure \ref{fig:layer_gamut} shows color gamuts computed using $L=3,6,12,18$ with the gamut volume as a fraction of the volume of sRGB. We can see how the volume increases with the number of layers. However, we found that the black point of the $18$-layer profile, $L^{*}=32.14$, is close to the minimum black point achievable with current CMYW materials--$L^{*}=31.55$ for a $3\cm$ cube of full CMY mixture. Thus, little additional gamut is likely to be gained by more layers.

Voxel layer $\ell$ is then defined as the set of voxels between isosurfaces $\ell$ and $\ell+1$, 
\begin{equation}
\label{eqn:layer_defn}
\voxelLayer{\ell} = \{ \vecv\in\voxelSet\ : \ d_{\ell} \leq d(\vecv)\ \mbox{and}\ \exists\ \vecu\in\window(\vecv)\ : d(\vecu)<d_{\ell} \}
\end{equation}
where the second condition is to ensure that the layers form thin voxel approximations of isosurfaces as much as possible, and $\window$ is the voxel window defined in Section \ref{sec:traversal:algorithm}. Due to different resolutions along each axis, $d_{\ell+1}-d_{\ell}$ may be multiple voxels thick depending on the orientation. In the case of $\voxelLayer{0}$, this condition is replaced by $\exists\ \vecu\in\window(\vecv)\ \suchthat\ \vecu\notin\voxelSet$.

Note that we can very efficiently compute normals for all isosurfaces, required for traversal of the error diffusion filter (see Section \ref{sec:traversal:algorithm}): $\vecn(\vecv)=\nabla\tilde{d}(\vecv)$, $\vecv\in\voxelLayer{\ell}$, where 
\begin{equation}
\label{eqn:signed_dist}
\tilde{d}(\vecu) = 
\begin{cases}
-d(\vecu) & \mbox{if } \vecu\in\voxelSet \\
d(\vecu)  & \otherwise
\end{cases}
\end{equation}
is a signed distance field, for $\vecu\in\boundingVoxels$. We use finite differences to compute $\nabla\tilde{d}$.

\subsection{Error Diffusion and Material Assignment}
\label{sec:layered:assign}

We treat each voxel layer as a set of surface voxels, traverse them as described in Section \ref{sec:traversal}, and halftone them as follows. 
Each tonal channel is halftoned independently using a standard error diffusion filter, mapped onto $\voxelLayer{\ell}$ as described in Section \ref{sec:traversal:mapping} and in the Appendix. At a given voxel $\vecv\in\voxelLayer{\ell}$, this results in a halftone vector $\sigHalftone(\vecv)\in\{0,1\}^T$ where $T = \mbox{dim}(\TonalValueSpace)$. Thus, multiple materials may have a tonal value of $1$ indicating they should be assigned to $\vecv$. Since we can only assign a single material $\matAssign(\vecv)$ to each voxel, we apply the following tie-breaking scheme.

For each tonal channel $t\in \{1,\dots,T\}$, we maintain a tie-breaker error, $c(t)$, which is initialized to $c(t)=0$ for all $t$ at the start of each slice. If at a given voxel, one or more tonal values are halftoned to $1$, the tonal channel $t^{*}=\arg\max_{t} c(t)$ with the largest tie-breaker error is declared the ``winner'' and we assign the corresponding material to the voxel. We then reset $c(t^{*})=0$ and increment $c(t)$ for all other $t$. If there is no $t$ such that $\sigHalftone(\vecv)[t]=1$, we assign white material to the voxel. 

Note that this tie-breaking scheme is independent of the per-tonal error diffusion, and the error diffusion does not know when a tonal is quantized to $1$, but the material is not assigned. This in general leads to a slight loss of tone, but is mitigated by the large number of voxels, and it is anyway accounted for by the color management, as described in Section \ref{sec:ColorManagement}, because we characterize the printer using targets printed with the same algorithm.

Note that errors are only diffused within the same layer, and each layer is halftoned independently. This has the following benefits. First, it allows different layers to be halftoned in parallel. Second, it further avoids the issue of resampling a halftoned signal, which would be necessary if the halftone was performed on the surface and then transfered to the interior. This is problematic because the halftoned signal is by design high-frequent and resampling it introduces artifacts. Third, we can use the threshold modulation of the halftoning algorithm~\cite{ZhouFang_varcoeff_threshmod_2003} to avoid inter-layer correlations, which can create dot gain artifacts (see Section \ref{sec:LightScatteringAbsorption}). We do this by seeding the pseudo-random number generators used for the threshold modulation differently for each slice and layer.

As discussed in Section \ref{sec:layered:construct}, due to differences in resolution along the different axes, some interior voxels will fall in between layers, and not be directly halftoned. For these voxels, we simply assign the material of the nearest voxel that does belong to a layer.

\section{Experiments}
\label{sec:experiments}

\subsection{Printing Setup}
We use an Objet500 Connex3 from Stratasys~\cite{Connex3}, for which the vendor has provided an interface allowing us to specify material assignments at the printer resolution. As the Connex3 allows only 3 model materials and 1 support material, we expanded the possible gamut by coloring the support material with a yellow acid dye. Thus, we print with white, cyan, magenta and yellow materials. Care must be taken that there are no large agglomerations of support within the object, which would weaken the structural integrity of the print. Fortunately, as our halftone is by design high-frequency, this is avoided by scaling the yellow tonal value by a constant factor between $0$ and $1$. We found $0.3$ to provide both a chromatic yellow and structurally stable prints. Our layer-based approach also allows us to prevent support from being present in the outermost layer (the surface), by applying a layer and tonal dependent soft-thresholding to the tonal values. For error diffusion we use a tone-adaptive filter~\cite{ostromoukhov_errdiff_2001} and threshold modulation technique~\cite{ZhouFang_varcoeff_threshmod_2003}.

\subsection{Color Management} \label{sec:ColorManagement}
\begin{figure}
\includegraphics[width=0.23\textwidth]{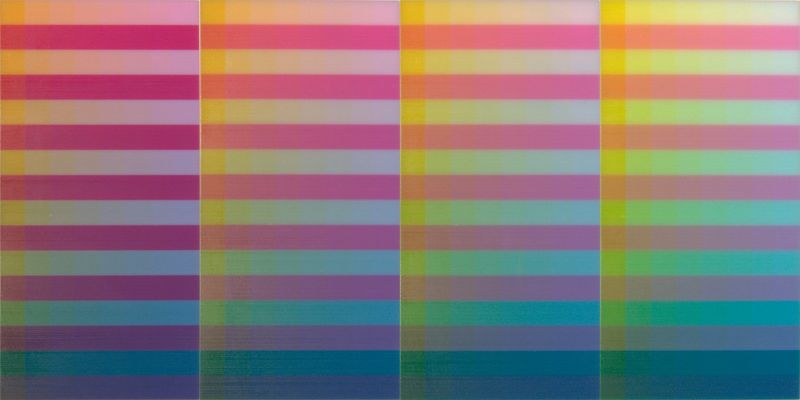}\hfill
\includegraphics[width=0.23\textwidth]{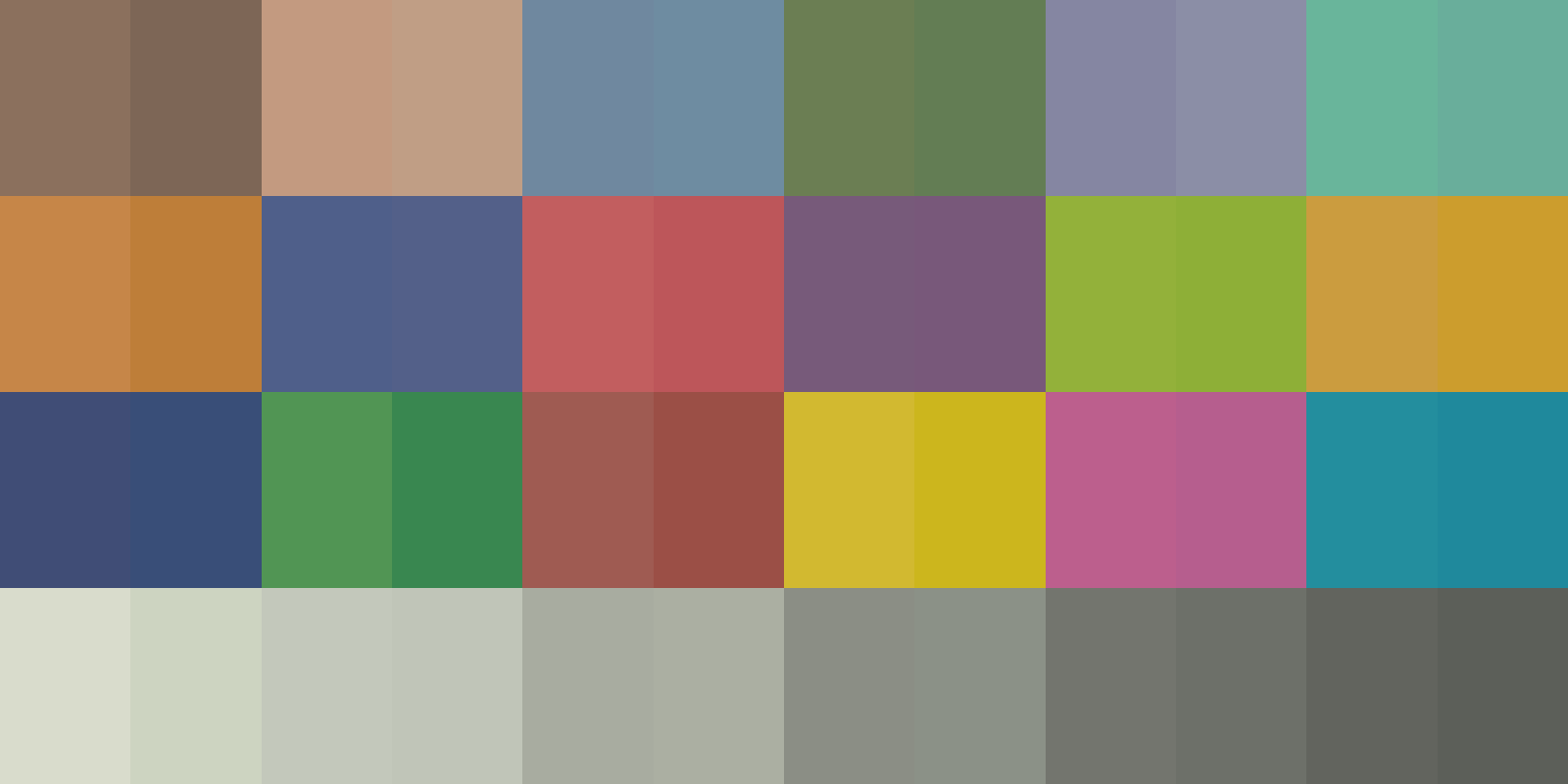}
\caption{Left: color characterization target with 12 layers, which was used to compute the ICC profile use to print the results in this paper. Right: predicted and measured color patches for a color checker; the left of each patch is the predicted color (gamut mapped) and the right is the measured color of the printed patch.}
\label{fig:color_manage}
\end{figure}

Specifying the color-to-tonal transformation $\sigColorTonal$ requires at least two ICC profiles: An input profile that specifies texture-data colorimetrically and a printer profile that includes gamut mapping and color separation. Generating a printer profile involves a colorimetric characterization of the printer, i.e. we need to predict the printout's CIEXYZ color for each tonal value vector in \TonalValueSpace. For this, the tonal value space is sampled and the color of corresponding printouts is measured and used to fit a colorimetric printer model. We used a fully empirical model, which interpolates between color-measured printouts of densely sampled tonal values--for our CMYW printer we use a  uniform $8\times8\times8$ sampling of the channel-wise linearized CMY tonal value cube. Linearization is performed employing the broadband Murray-Davies model according to~\cite{WybleBerns2000}. This target is printed using the algorithm described in Sections \ref{sec:traversal} and \ref{sec:layered}--exactly the same algorithm as used to generate the results shown in Figures \ref{fig:visual:objects} and \ref{fig:gallery}, without color management as the input are direct tonal values. A picture of the resulting printout is shown in Figure \ref{fig:color_manage} (left).

Measuring colors of such highly translucent printing materials is a challenging problem. It can be shown that measurements made by spectrophotometers used in the graphic arts community are systematically biased towards lower reflectance due to subsurface light transport away from the detection area. For our color measurements, we used a bidirectional $0^{\circ}/45^{\circ}$ measurement geometry, an almost colorimetric DSLR camera (Vora value of 0.9455~\cite{VoraTrussell1993}), diffuse broadband illumination simulating CIED50, flat fielding (to account for optical path variations between light source, printed color patches and camera), and a polynomial approach to map camera RGB values to CIEXYZ. Color errors determined by spectroradiometric measurements are within the interinstrument-variability of spectrophotometers used in graphic arts for opaque materials. For more details, the reader is referred to~\cite{OurPaper2015}.

We evaluated the effectiveness of color management as follows. We took a standard color checker and measured each of the 24 patches with a spectrophotometer. We then mapped each of these colors into our printer gamut (12-layers, Figure \ref{fig:layer_gamut}, second from right), and printed planar patches. We then measured the printed patches with the almost colorimetric DSLR camera, and computed the CIEDE2000~\cite{CIEDE2000} differences to the gamut mapped colors--i.e. the colors predicted by our empirical printer model. We observed a median error of $2.2$, a mean error of $2.3$, and minimum and maximum errors of $0.5$ and $6.1$. The predicted and printed colors are shown in Figure \ref{fig:color_manage} (right). Note that, even though the proposed half-toning algorithm was used to print the planar patches, this evaluates \emph{only the empirical printer model} and not the halftone, which is evaluated in Sections \ref{sec:experiments:eval} and~\ref{sec:experiments:visual}.

\subsection{Evaluation}
\label{sec:experiments:eval}
\qheading{Performance.}
Our non-optimized implementation supports multi-threading for some computations (for example the transfer operator and halftoning each layer in parallel), but does not exploit any GPU capabilities, although several aspects would lend themselves well to GPU implementation. Table \ref{tab:performance} shows performance data for several prints: the number of voxels, both those assigned a material ($|\voxelSet|$) and in the build volume ($|\boundingVoxels|$), computation times include the time to deliver the first chunk of slices to the printer and the total, and the print times. The results were collected on a standard desktop PC with an Intel Core i7-4770 processor and $32$ GB of memory.

\begin{table*}
\tbl{Performance details of our method.}{
\centering
\begin{tabular}{c|cc|cc|c|cccc}
Model & \multicolumn{2}{c|}{Voxel Count} & \multicolumn{2}{c|}{Compute Time} & Print Time & \multicolumn{4}{c}{Material-Tonal RMSE} \\
 & $|\voxelSet|$ & $|\boundingVoxels|$ & first chunk & total & & C & M & Y & W \\
\hline 
\parbox{0.15\textwidth}{\centering Bald head ($20\cm$)\\Figure \ref{fig:gallery}d} & $8.6\times10^{9}$ & $2.49\times10^{10}$ & 1 min & 6 h 10 min & 21 h & $0.0038$& $0.0074$ & $0.0074$ & $0.0184$ \\
\hline
\parbox{0.15\textwidth}{\centering Ruslan ($15\cm$)\\Figure \ref{fig:gallery}a} & $5.7\times10^{9}$ & $1.36\times10^{10}$ & 1 min & 4 h 18 min &  11 h$^*$ & $0.0097$& $0.0038$ & $0.0107$ & $0.0236$ \\
\hline
\parbox{0.15\textwidth}{\centering Ruslan ($5\cm$)\\Figure \ref{fig:gallery}b} & $2.12\times10^{8}$ & $5.07\times10^{8}$ & 8 s & 16 min &  2 h & $0.0094$& $0.0041$ & $0.0120$ & $0.0243$ \\
\hline
\parbox{0.15\textwidth}{\centering Nefertiti ($20\cm$)\\Figure \ref{fig:gallery}e} & $5.0\times10^{9}$ & $1.67\times10^{10}$ & 1 min & 4 h 55 min & 20 h & $0.0056$& $0.0041$ & $0.0080$ & $0.0173$ \\
\hline 
\parbox{0.15\textwidth}{\centering Apple ($8\cm$)\\Figure \ref{fig:gallery}c} & $1.8\times10^{9}$ & $3.7\times10^{9}$ & 30 s & 1 h & 10 h & $0.0056$& $0.0030$ & $0.0308$ & $0.0247$
\end{tabular}
}
\begin{tabnote}
\centering
See text for a description of how we compute the tonal errors. $^*$Print job terminated after $69\%$ of slices.
\end{tabnote}
\label{tab:performance}
\end{table*}

\qheading{Tone Preservation.}
It is in general difficult to evaluate how well a 3D print reproduces the color of the input texture, due to the large change in appearance resulting from different illuminants and illumination distribution in combination with the curved surface. As discussed further in Section \ref{sec:experiments:limits}, the printing materials we use are highly color inconstant, meaning the perceived color changes under different illuminants. Further, typically the textured model is not rendered using the reflectance properties of the printing materials. In the case of scanned objects, evaluating with respect to the original object is meaningless, since the capture systems are typically not color calibrated. 

Therefore, we evaluate the quality of tone preservation, rather than the similarity of perceived colors, as follows. We compute the root mean squared error (RMSE) between material assignments $m$ and tonal values $\sigTonalTrans$, which are given for several models in the last column of Table \ref{tab:performance}. First, we compute material fractions for each slice--the number of voxels assigned a given material in the slice divided by the total number of voxels $\vecv\in\voxelSet(s)$ such that $d(\vecv)<\maxdist$. Then we determine expected material fractions from the tonal values $\sigTonalTrans(\voxelSet(s))$ using the Demichel equations~\cite{Demichel1924a}. For example, we compute the expected material fractions for cyan and white from tonal values as 
\begin{eqnarray}
E_{C} & = C(1-M)(1-Y) +\frac{1}{2}CM(1-Y)\nonumber\\ & +\frac{1}{2}C(1-M)Y + \frac{1}{3}CMY
\end{eqnarray}
and
\begin{equation}
E_{W} = (1-C)(1-M)(1-Y)
\end{equation}
respectively, where $C$, $M$ and $Y$ are the average tonal values over the same set of voxels. The formulas are similar for magenta and yellow.
Finally, we compute the errors as the difference between the expected material fraction and the actual material fraction for each tonal channel. We see that our algorithm preserves tone very well, with most errors being less than $1\%$ of the tonal value range.

\subsection{Visual Experiment}\label{sec:experiments:visual}
\begin{figure}[!b]
\includegraphics[width=0.48\textwidth]{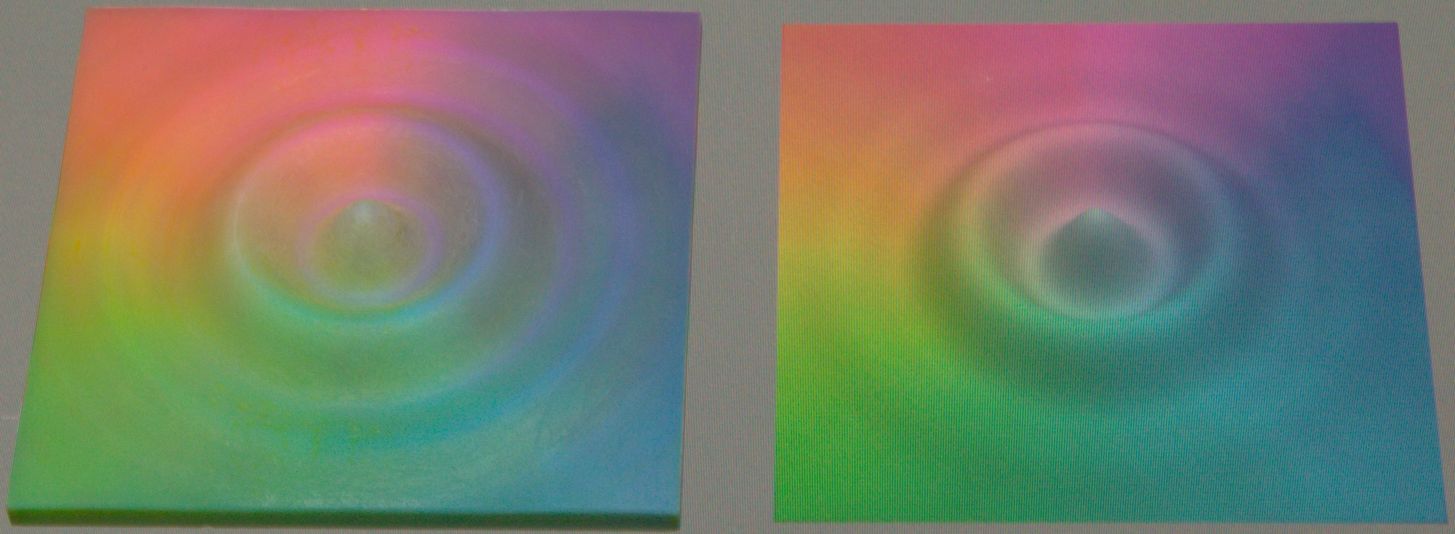}\\
\includegraphics[width=0.48\textwidth]{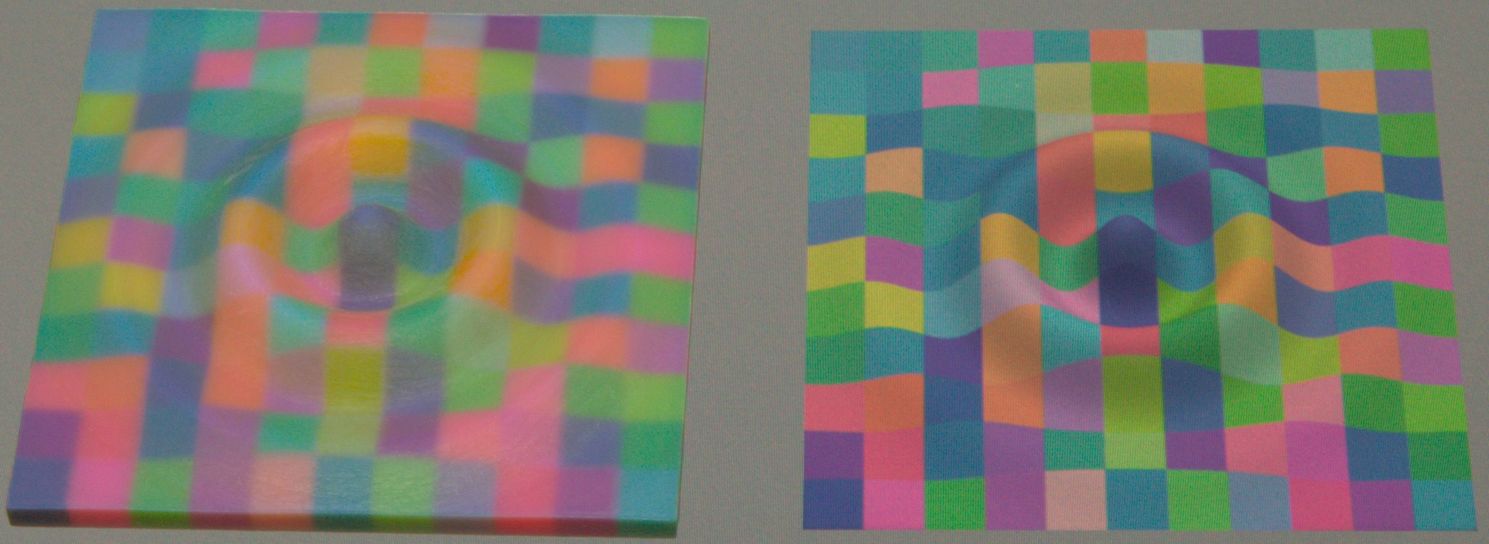}
\caption{Pictures of the test surfaces taken from the chin-rest position in our experiment; printed (left) and rendered (right). Moir\'{e} on the renderings is caused by the acquisition process.}
\label{fig:visual:objects}
\end{figure}

Visual experiments were conducted to assess the structural quality of the proposed halftone method, in particular the level of graininess and the visibility of structural artifacts. For this, we generated a test surface with the following function
\begin{equation}
z(x,y) = e^{-r^2/\sigma^2} (0.5\cos(3r)+0.5)
\end{equation}
where $r=\sqrt{x^2+y^2}$, $\sigma=2R/3$, $R=5\cm$ with both $x$ and $y$ ranging from $-5\cm$ to $5\cm$. We then applied two textures to this surface, as shown in Figure \ref{fig:visual:objects}; one with smooth color gradients and one with high-frequency patches; the colors were selected to uniformly cover the a*b*-plane of lightness L*=70 and gamut-mapped using the absolute-colorimetric intent of our ICC profile.

The printed surfaces were placed on a colorimetrically characterized display next to their rendered versions (gap of approx. 1.25 cm - the left/right position of rendering and print was randomly chosen for each subject). We used an \emph{Eizo ColorEdge cg301w} $30$-inch display with a resolution of $2560\times 1600$ pixels. The printed surfaces were $10\cm \times 10\cm$, which covers $400\times 400$ pixels on the screen. The screen was placed horizontally and illuminated from directly above using a \emph{JUST NORMLICHT LED Color Viewing Light M} viewing booth at a distance of $1\m$ from the screen to provide a diffuse illumination. The illuminant used was LED-simulated CIED50. A chin-rest was placed so to obtain $60$ pixels-per-degree (ppd) at the screen center ($85\cm$ distance) and a $45^{\circ}$ viewing angle to the screen. The renderings were done to approximately match this viewing condition and the luminance level of the print. Figure \ref{fig:visual:objects} shows pictures taken with an almost colorimetric DSLR camera of what the subjects viewed. A detailed image and diagram of the setup can be found in the supplemental material.

30 naive subjects with normal visual acuity -- tested with the Snellen test -- participated in the experiments (10 female and 20 male; average age of 30 years). For each of the two textures, a total of 10 distorted textures were created by adding zero-mean Gaussian pixel-noise with standard deviations of $1,\dots, 10$ CIEDE2000 units, simulating different levels of graininess. In a first experiment, subjects were asked to select the rendering with the distorted texture matching the graininess of the print. In a second experiment, they had to judge the level of structural artifacts in the print -- showing the noise-free rendering as reference -- on a quality scale of 0-5: 0 (not visible), 1 (visible but not disturbing), 2 (slightly disturbing), 3 (disturbing), 4 (very disturbing), 5 (extremely disturbing).

\begin{table}
\tbl{Results of the visual Experiments.}{%
\centering
\begin{tabular}{r|cccc}
\multicolumn{5}{c}{Experiment I (graininess level [noise std. in CIEDE2000])}\\ \hline
							& mean & std & min & max \\
Smooth texture & 0.9667 & 0.9994 & 0 & 4\\
Patch texture & 2.0000  &   1.2594    &      0  &   4 \\ \hline
\multicolumn{5}{c}{Experiment II (structural artifacts [quality scale 0-5])}\\ \hline
              & mean & std & min & max \\
Smooth texture & 1.3000  &  0.7497 & 0 & 3  \\
Patch texture & 1.3667  &  0.8899  & 0 & 4 \\ \hline
\end{tabular}
}
\label{tab:experiment}
\end{table}

Results of the experiments are shown in Table \ref{tab:experiment}. For both textures, the average perceived graininess level corresponds to a noise standard deviation smaller or equal to 2 CIEDE2000 units. To put this value into perspective, we computed the pixel-wise CIEDE2000 differences between the noise-free and distorted texture (noise std. of 2 CIEDE2000 units) using S-CIELAB~\cite{ZhangWandell1996} with a visual resolution of 60 ppd~\cite{JohnsonFairchild2003}. The 99th percentile of CIEDE2000 errors is 1.644 (smooth texture) and 1.694 (patch texture). Since this is only slightly above the just noticeable distance, we can conclude that the perceived graininess of the halftone is very low. The second experiment revealed that also structural artifacts do not adversely affect the perceived quality of the halftone: for both textures, average quality scores are smaller than 1.4, i.e. artifacts are visible but not even slightly disturbing. Note that there are some drying related artifacts, which are likely considered in the quality scores but independent of the halftone.\\

\subsection{Qualitative Results}

\begin{figure*}[!ht]
\centering
\parbox{9cm}{%
\centering
\framebox[9cm]{%
\parbox{2.63cm}{\centering\includegraphics[height=4.5cm]{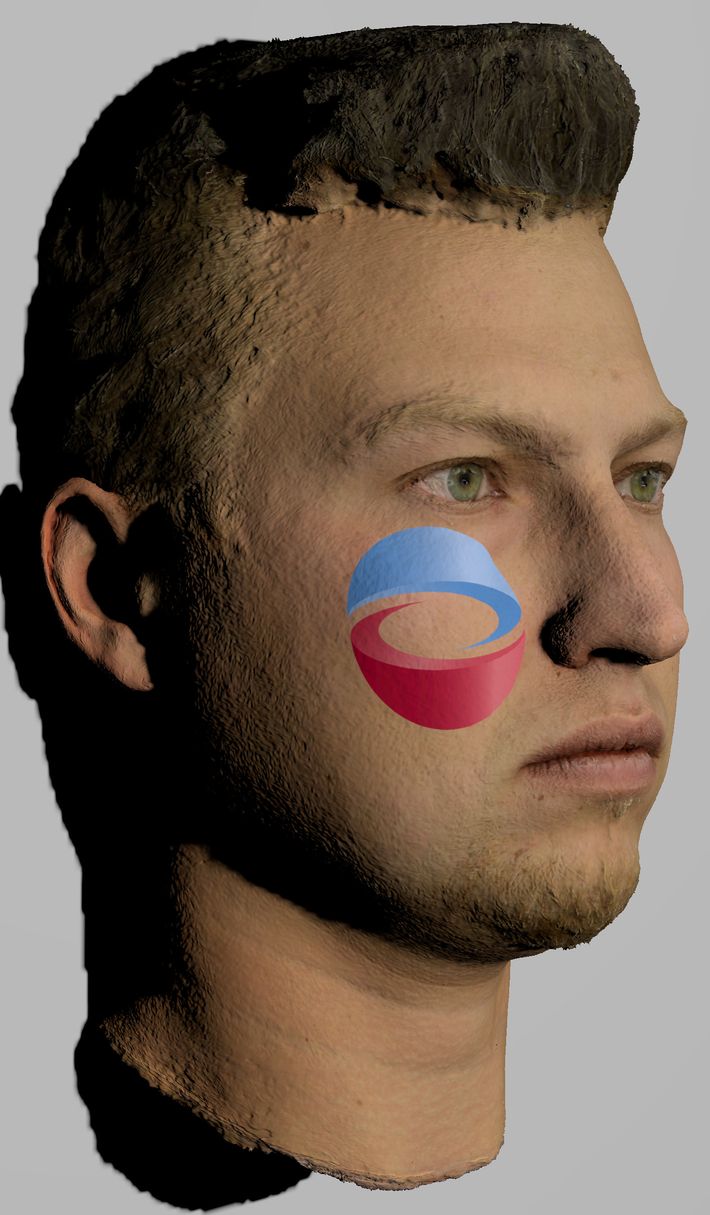}\\original texture}\hfill
\parbox{2.63cm}{\centering\includegraphics[height=4.5cm]{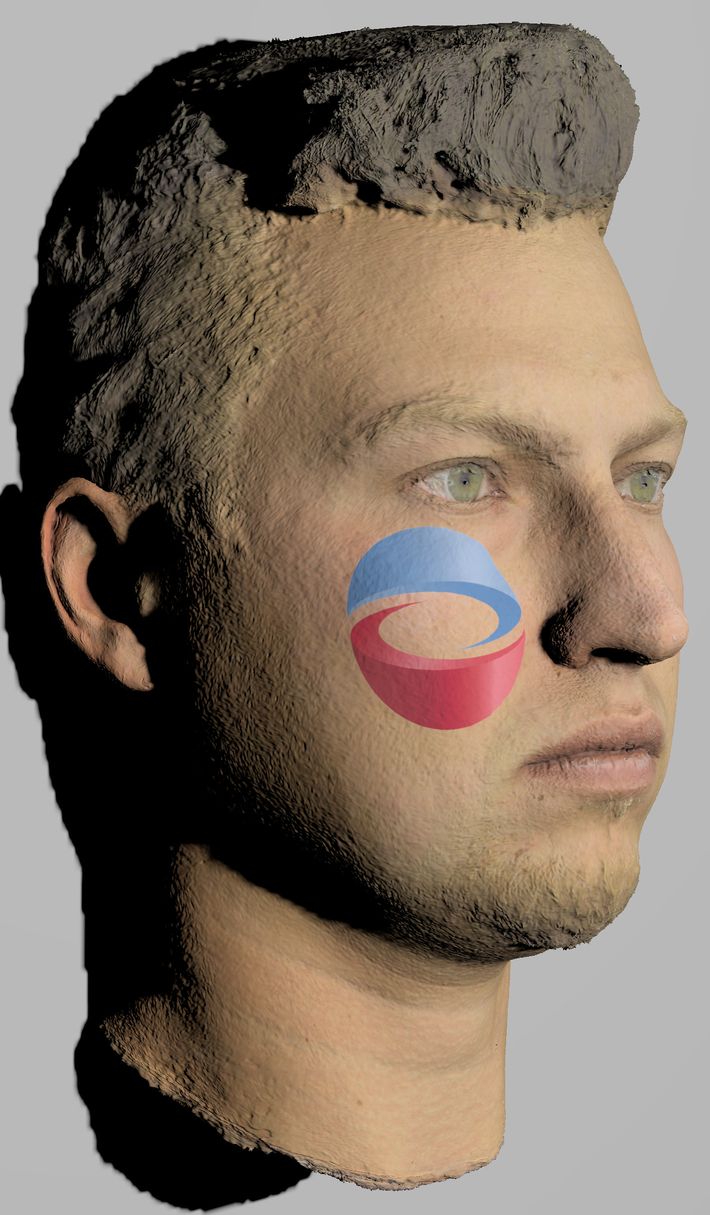}\\simulated texture}\hfill
\parbox{2.82cm}{\centering\includegraphics[height=4.5cm]{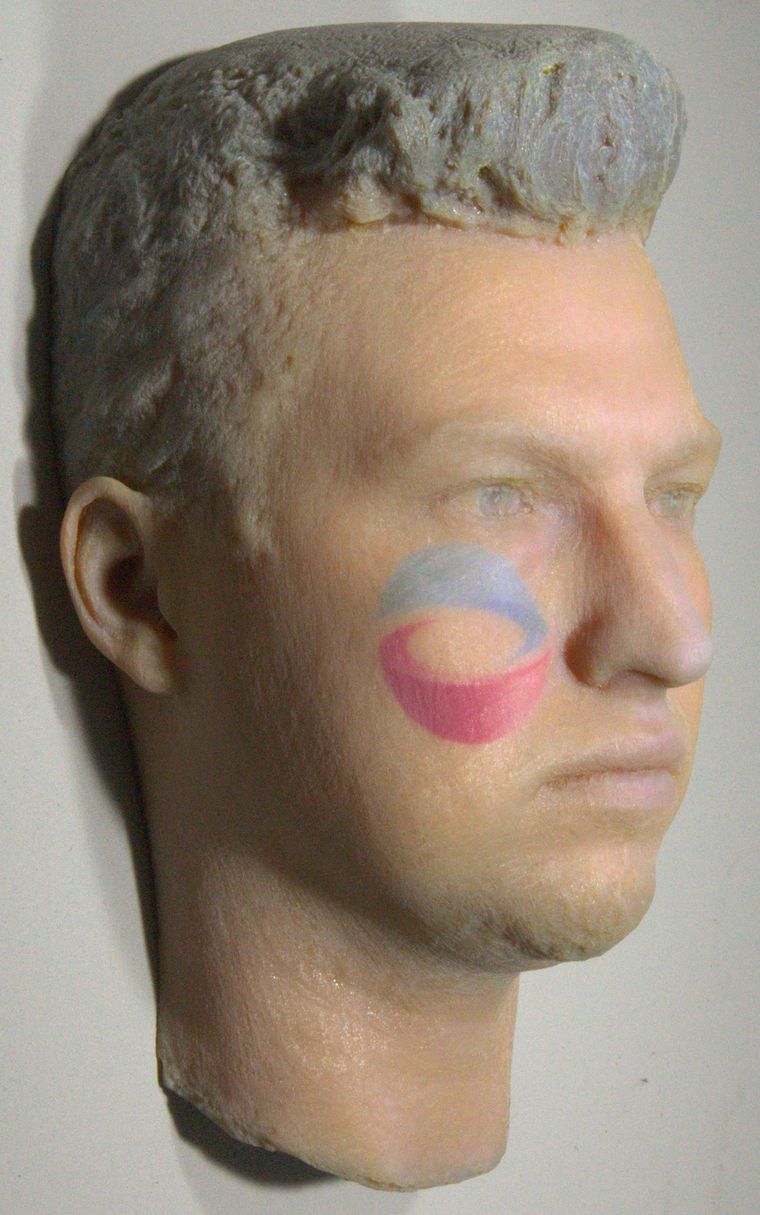}\\$15\cm$ print}%
}\\
(a)
}
\hfill
\parbox{4.5cm}{
\centering
\fbox{%
\parbox{3.86cm}{\centering\includegraphics[height=4.5cm]{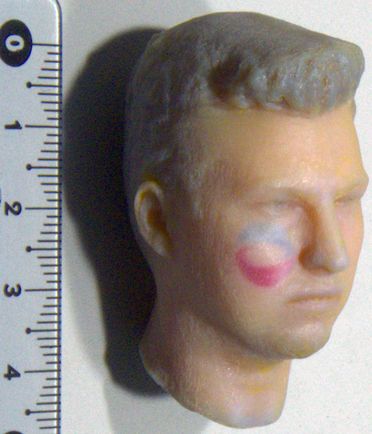}\\$5\cm$ print}%
}\\
(b)
}
\hfill
\parbox{4.0cm}{
\centering
\fbox{%
\parbox{3.73cm}{\centering\includegraphics[height=4.5cm]{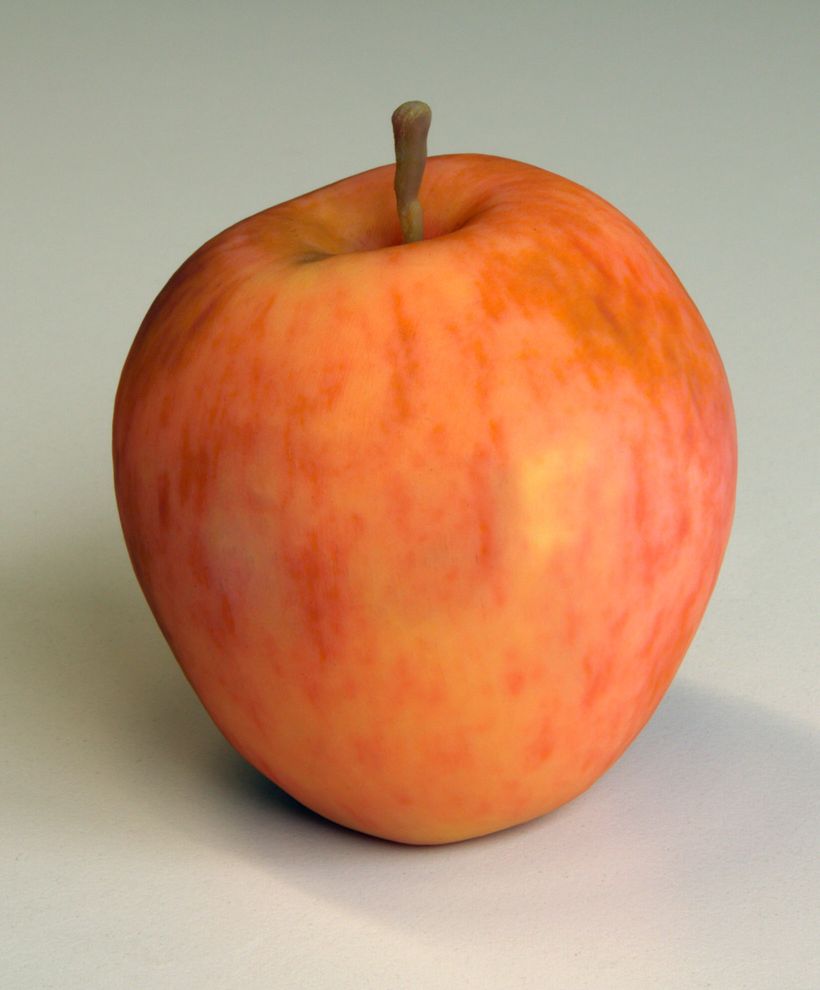}\\$8\cm$ printed apple}%
}\\
(c)
}
\\
\vspace{0.2cm}
\parbox{8.0cm}{
\centering
\framebox[8.0cm]{%
\parbox{3.98cm}{\centering\includegraphics[height=4.5cm]{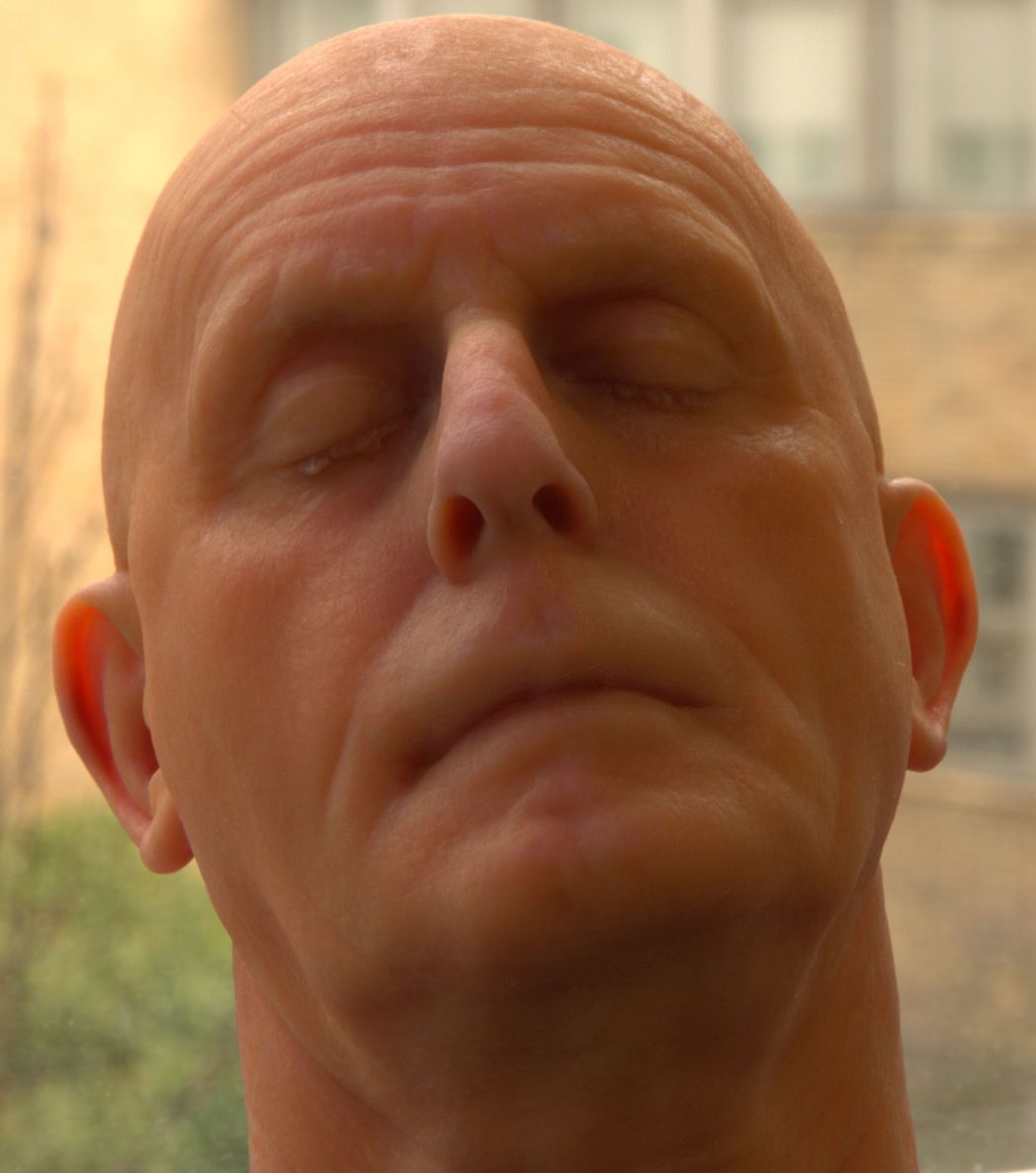}\\back-lit $20\cm$ print}\hfill%
\parbox{1.80cm}{\centering\includegraphics[height=4.5cm]{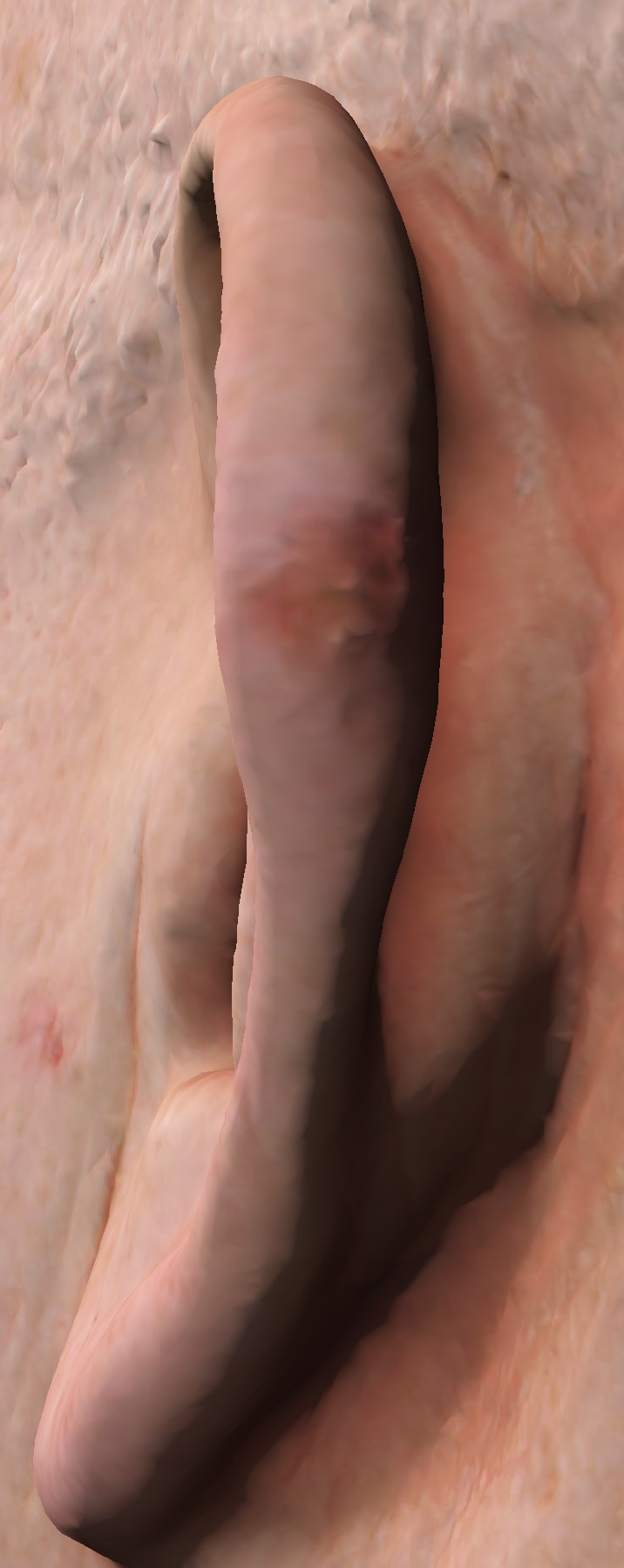}\\rendered ear}\hfill%
\parbox{1.66cm}{\centering\includegraphics[height=4.5cm]{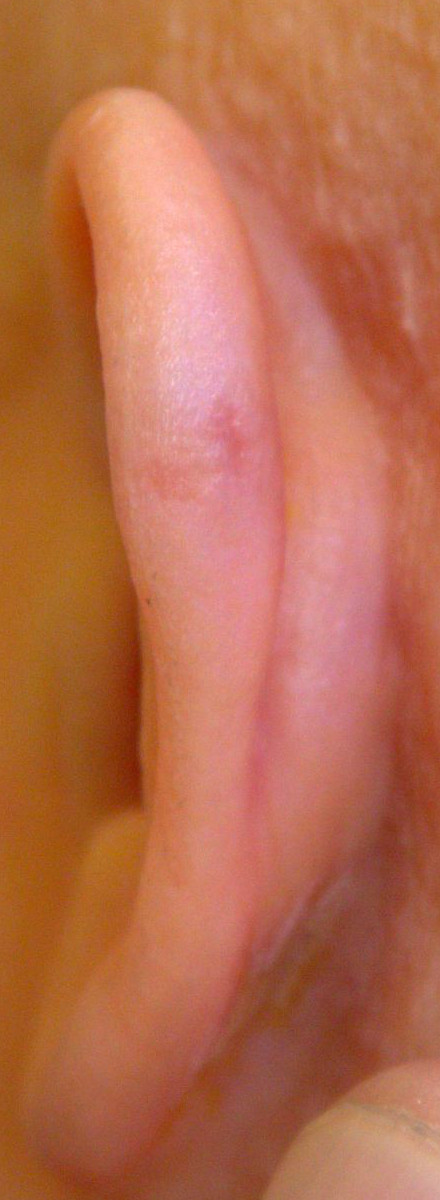}\\printed ear}%
}\\
(d)
}
\hfill
\parbox{8.0cm}{
\centering
\framebox[8.0cm]{%
\parbox{2.25cm}{\centering\includegraphics[height=4.5cm]{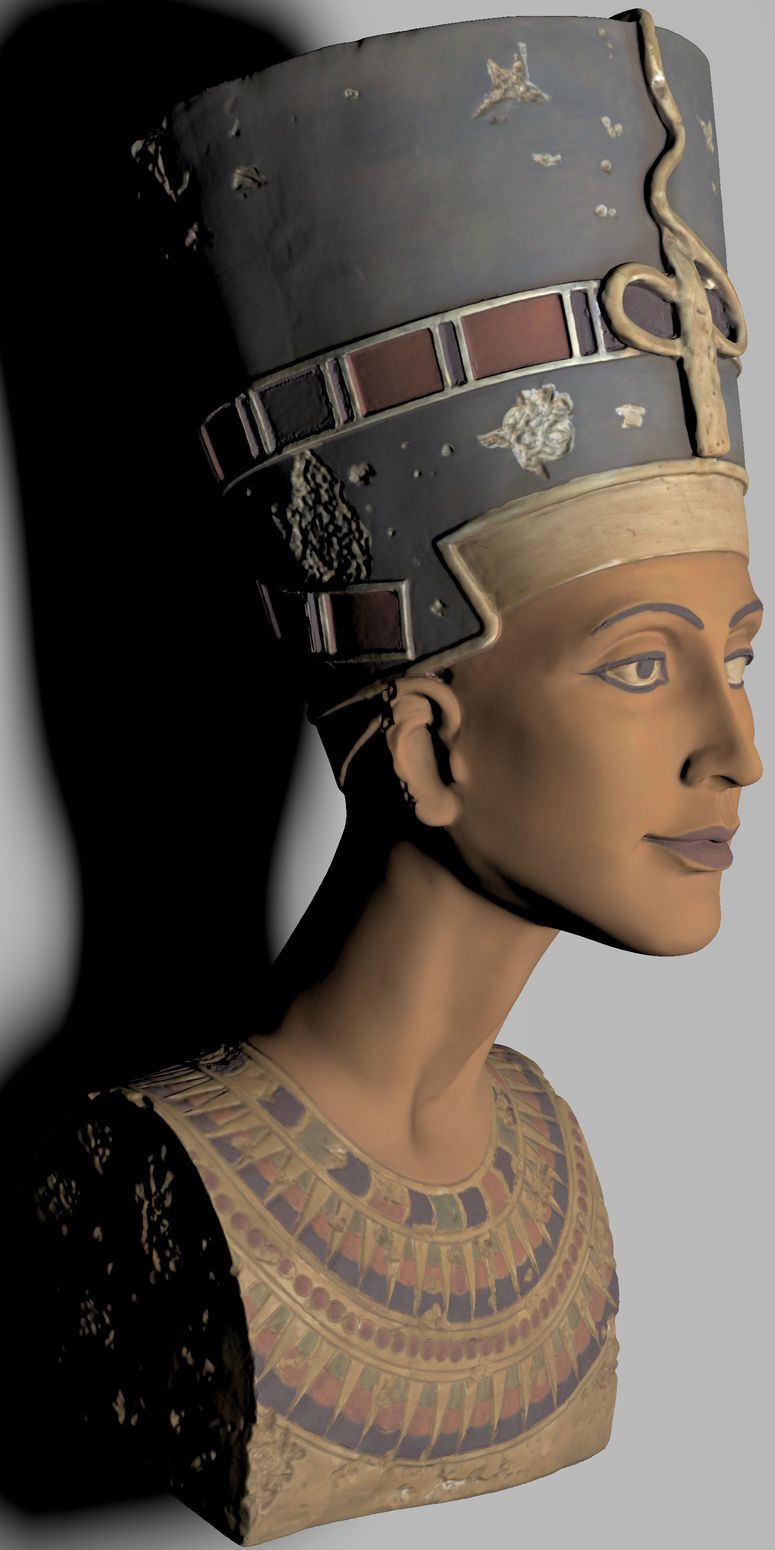}\\rendering}\hfill%
\parbox{2.25cm}{\centering\includegraphics[height=4.5cm]{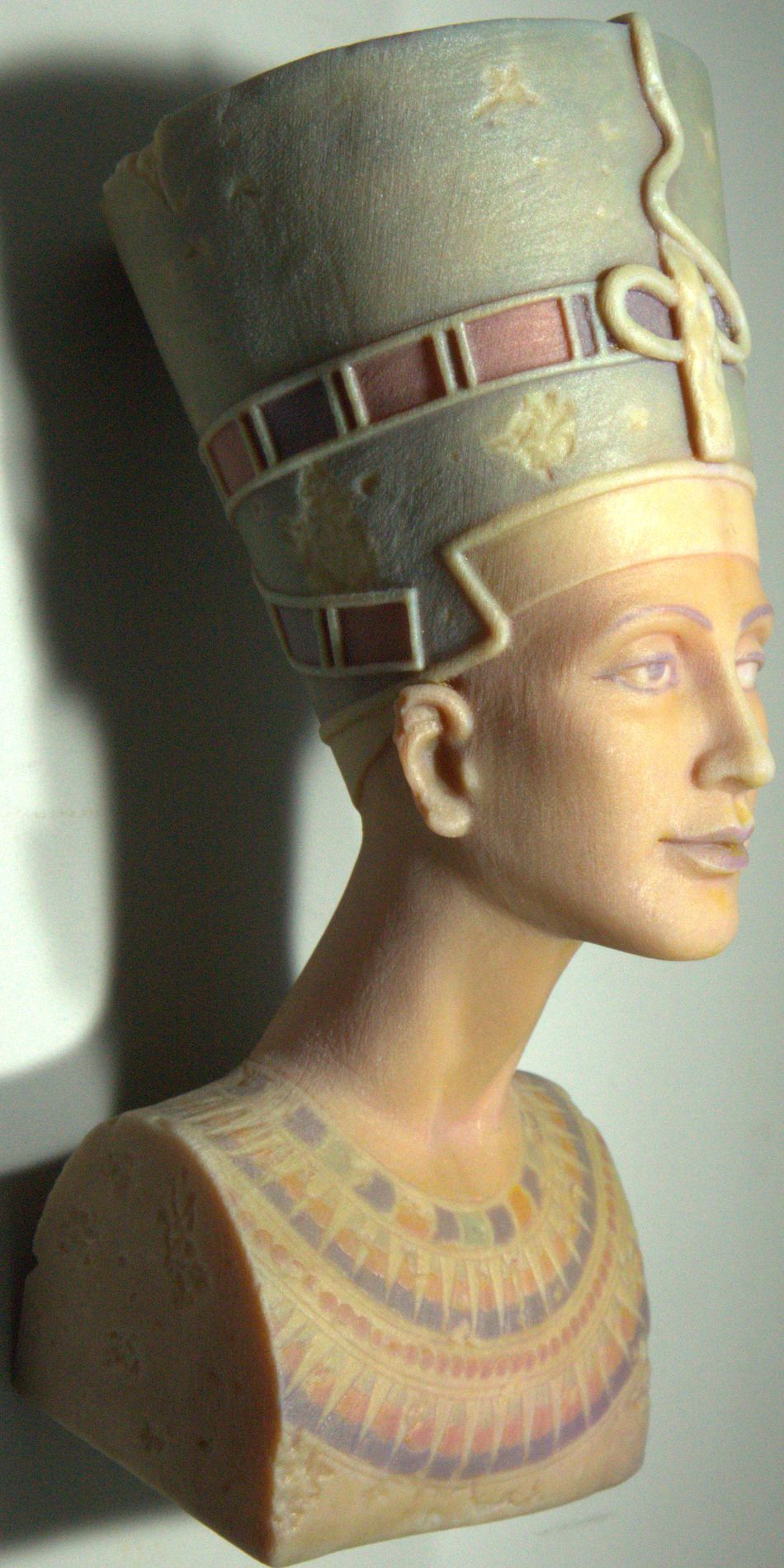}\\$20\cm$ print}\hfill%
\parbox{2.56cm}{\centering\includegraphics[height=4.5cm]{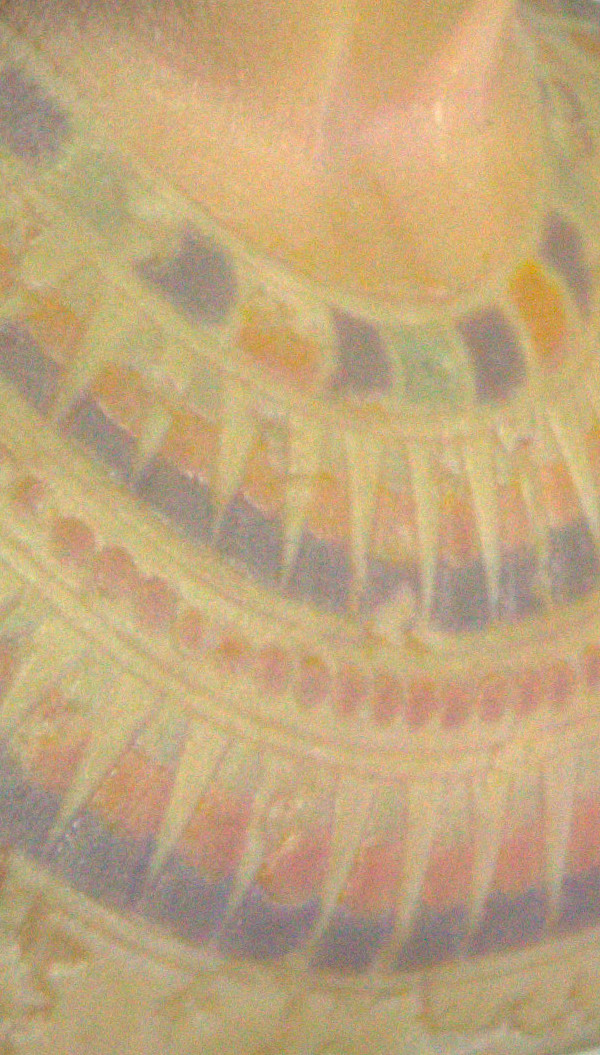}\\zoomed in}%
}\\
(e)
}
\caption{Various 3D color prints generated using our software. Note that color deviations between rendering and print may have various reasons such as gamut limitations, goniochromatism and color inconstancy of printing materials, light field and illuminant differences between real scene and rendering, etc. The comparison aims to show how well texture details are preserved by our halftoning approach and that no artifacts are introduced. See the text for discussion of the individual parts.}
\label{fig:gallery}
\end{figure*}

Figure \ref{fig:gallery} shows some qualitative results demonstrating the detail and realism produced by our software using our hardware and material setup. Part (a) shows (left-to-right) a head scan~\cite{Ruslan_TurboSquid} with original texture rendered with diffuse shading, the same model with the texture gamut-mapped using our profile, and the resulting $3$D print ($15\cm$ tall). Note how well both smooth tones and high-frequency details are reproduced. Part (b) shows the same model printed $5\cm$ tall to show the effects of scale on the prints--note how similar the features of the two prints are despite the difference in scale. Part (c) shows a realistic-looking printed apple~\cite{Apple_TurboSquid}. Parts (d) and (e) demonstrate how well our halftoning algorithms reproduce high-frequency details in the textures. Prints (c)-(e) were generated with an error in the ICC profile, so the colors are off in absolute terms when compared to the textures. In part (d)~\cite{Head_Ten24}, the back of the ear is rendered next to the printed ear, which is about $3.5\cm$ high. The images of the prints in parts (a) and (e) were taken under simulated CIED50, to which we color-characterized the camera. We transformed the raw RGB images to CIEXYZ values as described in~\cite[Sec. 3.2]{OurPaper2015}, and transformed these values to sRGB for display. All other photographs of prints were taken using automatic camera settings.

\subsection{Limitations}
\label{sec:experiments:limits}
Our setup is limited primarily from the hardware side. With four materials we are forced to print without a black. This means we have difficulty reproducing dark colors and a lack of color constancy--a change of illuminant alters the perceived color. We measured the color inconstancy index (CII) between CIED50 and CIEF11 of cyan, magenta, yellow and white, to be $4.65$, $3.24$, $5.28$ and $0.59$, respectively, using CAT02~\cite{CIECAM022004} for chromatic adaptation and CIEDE2000 for color difference~\cite{Fairchild2013}. We also have no control over the translucency, as the materials have very similar scattering and transmission properties. We can only use their translucency to create realistic looking objects, not control it to a desired level. The translucency of the materials also results in blurring of some details; an interesting approach has been proposed to address this in a way similar to unsharp masking~\cite{Cignoni_2008}. The introduction of printers with more materials, and new, more opaque materials would address these limitations, and open the possibility of combining color with desired scattering and reflective properties. A compelling open question is whether such complex appearance properties can be achieved using error diffusion-style techniques. A further limitation is due to the colored support material. During the uv-curing process, support material adjacent to model material mixes and binds with the model material. This results in a yellow tinge on some sloping and vertical surfaces. This could be overcome without additional materials by a multi-pass printing, in which the support material is cured separately from the other materials.

Algorithmically, our approach has the same limitations as $2$D error diffusion methods: for extremely low tonal values, we encounter mild start-up artifacts--materials are not placed entirely uniformly. We have found this only in extreme cases.

The curvature of the surface affects the transfer of tonal values to the interior layers when the mean absolute curvature $H$ approaches $1/d_w$, where $d_w$ is the depth of highest reflectance as discussed in Section \ref{sec:prelim}; equivalently as the radius of curvature approaches $d_w$. High negative mean curvature (convex) regions may darken and high-frequency texture features that coincide with high positive mean curvature (concave) regions will ``bloom". We have not found this to be a major problem, although such effects show up strongest in small prints, e.g. the nose of the small Ruslan print in Figure \ref{fig:gallery} (b). We plan to address curvature-dependent effects in future work, e.g. by adjusting the tonal values according to curvature, which could be estimated using the signed distance field $\tilde{d}$ in \eqref{eqn:signed_dist}. The surfaces used in the visual experiment also demonstrate the extent to which curvature influences the perceived color. These surfaces, shown in Figure \ref{fig:visual:objects}, contain both concave and convex regions with high curvature, and relatively flat regions on the periphery.

\section{Conclusions}
In this paper, we have presented algorithms for precise and efficient control of material placement in multi-jet $3$D printers for the purposes of halftoning--a critical ingredient in accurate color reproduction. We have presented a layered halftone, which accounts for the high-translucency of currently available color materials, and a traversal algorithm for voxel representations of surfaces, which allows us to map $2$D error diffusion filters onto a surface in a consistently oriented way. This allows us to leverage decades of knowledge from $2$D error diffusion research. Our algorithms fit within an efficient streaming architecture, preserve tone and do not produce artifacts. We have further shown how our algorithms seamlessly allow the printer's support material to be colored, thereby expanding the printer's color gamut while preserving structural stability. The $3$D color prints generated with our setup exhibit a high level of detail and realism. 

Due to the ability to arbitrarily configure materials with different optical properties through an object, multi-jet printing has tremendous potential for graphical $3$D printing in terms of reproducing complex appearance properties. By providing the first full color capabilities for these printers, we have provided a critical building block for graphical $3$D printing.

The introduction of more opaque color materials will allow our algorithms to print with larger color gamuts using fewer layers, resulting in a performance boost. Better exploiting parallelism in our computations is a key direction for future work as printer resolutions and build volumes increase. As the number of printing materials increase, it will be important to consider printer models in characterizing the printers and computing profiles.

\section*{Acknowledgements}

We give a big shout-out to Tejas Tanksale (TT) for his help in generating ICC profiles and conducting measurements. In addition, we thank: Christoph Godau for access to the equipment at the IDD of TU Darmstadt; Stratasys for providing the voxel-level printing interface to the printer; Pedro Santos and the CultLab3D of Fraunhofer IGD for making the Nefertiti buste available to us; Christian Zeh from Clariant Produkte GmbH for coloring the support material; Ten24 for making the head model available, and ScanLab and TurboSquid for making the Ruslan scan available. This work was funded by the FhG Internal Programs under Grant No. Attract 008-600075.

\section*{Appendix}
\label{sec:appendix}
Our halftoning algorithms we propose are based on error diffusion, in which after a pixel or voxel is quantized, the resulting error is distributed to nearby pixels or voxels using an error diffusion filter. Thus, if a pixel with a value of $0.5$ is quantized to $0$, the error is distributed to its neighbors, making it more likely that they are quantized to $1$. Formally, we are given an input tonal signal $\sigTonal$ and an error signal $\veca$, which is initialized $\veca(\vecv)=\textbf{0}\ \forall\ \vecv$. The error diffused signal is defined as
\begin{equation}
\label{eqn:err_diffused_signal}
\sigErrDiff(\vecv)=\sigTonal(\vecv)+\veca(\vecv), 
\end{equation}
which is then quantized using a threshold
\begin{equation}
\sigHalftone(\vecv) = 
	\begin{cases}
		1 & \mbox{if } \sigErrDiff(\vecv) > t \\
		0 & \otherwise \\
	\end{cases}
\end{equation}
the resulting error is then diffused to the neighbors
\begin{equation}
\veca(\vecu) = \veca(\vecu) - w_{\vecv,\vecu} \left( \sigHalftone(\vecv) - \sigErrDiff(\vecv) \right)
\end{equation}
where $\vecu\in\nbrs(\vecv)$ is a neighbor of $\vecv$, $w_{\vecv,\vecu}$ is an element of the error diffusion filter diffusing error from $\vecv$ to $\vecu$.

\bibliographystyle{acmtog}

\begin{thebibliography}{}

\bibitem[\protect\citeauthoryear{{3DS}ystems}{{3DS}ystems}{2014}]{Projet860}
{\sc {3DS}ystems}. 2014.
\newblock Projet {860Pro}.
\newblock http://www.3dsystems.com/3d-printers/professional/projet-860pro.

\bibitem[\protect\citeauthoryear{Agar and Allebach}{Agar and
  Allebach}{2005}]{AgarAllebach2005}
{\sc Agar, A.} {\sc and} {\sc Allebach, J.} 2005.
\newblock Model-based color halftoning using direct binary search.
\newblock {\em {IEEE} Trans. on Image Proc.\/}~{\em 14,\/}~12, 1945--1959.

\bibitem[\protect\citeauthoryear{Alexa and Kyprianidis}{Alexa and
  Kyprianidis}{2015}]{Alexa2015}
{\sc Alexa, M.} {\sc and} {\sc Kyprianidis, J.} 2015.
\newblock Error diffusion on meshes.
\newblock {\em Computers and Graphics (Proc. SMI 2014)\/}~{\em 46}, 336--344.

\bibitem[\protect\citeauthoryear{Arikan, Brunton, Tanksale, and Urban}{Arikan
  et~al\mbox{.}}{2015}]{OurPaper2015}
{\sc Arikan, C.}, {\sc Brunton, A.}, {\sc Tanksale, T.}, {\sc and} {\sc Urban,
  P.} 2015.
\newblock {Color-Managed 3D-Printing with highly Translucent Printing
  Materials}.
\newblock In {\em SPIE/IS\&T Electronic Imaging Conference (accepted --
  included in supplemental material)}. San Francisco.

\bibitem[\protect\citeauthoryear{Bayer}{Bayer}{1973}]{Bayer1973}
{\sc Bayer, B.~E.} 1973.
\newblock An optimum method for two-level rendition of continuous-tone
  pictures.
\newblock In {\em {IEEE} Intl. Conf. on Comm.} Seattle, WA, 11--15.

\bibitem[\protect\citeauthoryear{Campbell, Kulikowski, and Levinson}{Campbell
  et~al\mbox{.}}{1966}]{CampbellKulikowskiLevinson1966}
{\sc Campbell, F.}, {\sc Kulikowski, J.}, {\sc and} {\sc Levinson, J.} 1966.
\newblock The effect of orientation on the visual resolution of gratings.
\newblock {\em The Jour. of Physiology\/}~{\em 187,\/}~2, 427--436.

\bibitem[\protect\citeauthoryear{Chandrasekhar}{Chandrasekhar}{1960}]{Chandrasekhar1960}
{\sc Chandrasekhar, S.} 1960.
\newblock {\em Radiative transfer}.
\newblock Courier Dover Publications.

\bibitem[\protect\citeauthoryear{Chang, Alain, and Ostromoukhov}{Chang
  et~al\mbox{.}}{2009}]{Chang_structaware_ed_2009}
{\sc Chang, J.}, {\sc Alain, B.}, {\sc and} {\sc Ostromoukhov, V.} 2009.
\newblock Structure-aware error diffusion.
\newblock {\em {ACM} TOG (Proc. SIGGRAPH Asia)\/}~{\em 28,\/}~5, 162:1--162:8.

\bibitem[\protect\citeauthoryear{Chang and Allebach}{Chang and
  Allebach}{2003}]{Chang_memeffic_ed_2003}
{\sc Chang, T.} {\sc and} {\sc Allebach, J.} 2003.
\newblock Memory efficient error diffusion.
\newblock {\em {IEEE} Trans. on Image Proc.\/}~{\em 12,\/}~11, 1352--1366.

\bibitem[\protect\citeauthoryear{Chen, Levin, Didyk, Sitthi-Armorn, and
  Matusik}{Chen et~al\mbox{.}}{2013}]{Chen_siggraph2013}
{\sc Chen, D.}, {\sc Levin, D.}, {\sc Didyk, P.}, {\sc Sitthi-Armorn, P.}, {\sc
  and} {\sc Matusik, W.} 2013.
\newblock Spec2fab: A reducer-tuner model for translating specifications to
  {3D} prints.
\newblock {\em {ACM} TOG (Proc. SIGGRAPH)\/}~{\em 32,\/}~4.

\bibitem[\protect\citeauthoryear{Cho, Sachs, Patrikalakis, and Troxel}{Cho
  et~al\mbox{.}}{2003}]{ChoSachsPatrikalakisTroxel2003}
{\sc Cho, W.}, {\sc Sachs, E.}, {\sc Patrikalakis, N.~M.}, {\sc and} {\sc
  Troxel, D.~E.} 2003.
\newblock A dithering algorithm for local composition control with
  three-dimensional printing.
\newblock {\em CAD\/}~{\em 35,\/}~9, 851--867.

\bibitem[\protect\citeauthoryear{{CIE Publication No. 142}}{{CIE Publication
  No. 142}}{2001}]{CIEDE2000}
{\sc {CIE Publication No. 142}}. 2001.
\newblock {Improvement to Industrial Colour-Difference Evaluation}.
\newblock Tech. rep., {Central Bureau of the CIE}, {Vienna, Austria}.

\bibitem[\protect\citeauthoryear{{CIE Publication No. 159}}{{CIE Publication
  No. 159}}{2004}]{CIECAM022004}
{\sc {CIE Publication No. 159}}. 2004.
\newblock {A colour appearance model for colour management systems: CIECAM02}.
\newblock {CIE Central Bureau}, {Vienna, Austria}.

\bibitem[\protect\citeauthoryear{Cignoni, Gobbetti, Pintus, and
  Scopigno}{Cignoni et~al\mbox{.}}{2008}]{Cignoni_2008}
{\sc Cignoni, P.}, {\sc Gobbetti, E.}, {\sc Pintus, R.}, {\sc and} {\sc
  Scopigno, R.} 2008.
\newblock Color enhancement for rapid prototyping.
\newblock In {\em VAST}.

\bibitem[\protect\citeauthoryear{Demichel}{Demichel}{1924}]{Demichel1924a}
{\sc Demichel, E.} 1924.
\newblock Le proc{\'e}d{\'e}.
\newblock ~{\em 26,\/}~3, 17--21, 26--27.

\bibitem[\protect\citeauthoryear{Dong, Wang, Pellacini, Tong, and Guo}{Dong
  et~al\mbox{.}}{2010}]{Dong2010}
{\sc Dong, Y.}, {\sc Wang, J.}, {\sc Pellacini, F.}, {\sc Tong, X.}, {\sc and}
  {\sc Guo, B.} 2010.
\newblock Fabricating spatially-varying subsurface scattering.
\newblock {\em {ACM} TOG (Proc. SIGGRAPH)\/}~{\em 29,\/}~4.

\bibitem[\protect\citeauthoryear{Doubrovski, Tsai, Dikovsky, Geraedts, Herr,
  and Oxman}{Doubrovski et~al\mbox{.}}{2014}]{doubrovski2014}
{\sc Doubrovski, E.}, {\sc Tsai, E.}, {\sc Dikovsky, D.}, {\sc Geraedts, J.},
  {\sc Herr, H.}, {\sc and} {\sc Oxman, N.} 2014.
\newblock Voxel-based fabrication through material property mapping: A design
  method for bitmap printing.
\newblock {\em CAD\/}.

\bibitem[\protect\citeauthoryear{Eschbach and Knox}{Eschbach and
  Knox}{1991}]{EschbachKnox1991}
{\sc Eschbach, R.} {\sc and} {\sc Knox, K.} 1991.
\newblock Error-diffusion algorithm with edge enhancement.
\newblock {\em JOSA A\/}~{\em 8,\/}~12, 1844--1850.

\bibitem[\protect\citeauthoryear{Fairchild}{Fairchild}{2013}]{Fairchild2013}
{\sc Fairchild, M.~D.} 2013.
\newblock {\em Color Appearance Models\/}, 3 ed.
\newblock John Wiley \& Sons, inc., West Sussex, England.

\bibitem[\protect\citeauthoryear{Felzenzwalb and Huttenlocher}{Felzenzwalb and
  Huttenlocher}{2004}]{Felzenzwalb2004}
{\sc Felzenzwalb, P.} {\sc and} {\sc Huttenlocher, D.} 2004.
\newblock Distance transforms of sampled functions.
\newblock Tech. rep., Cornell University.

\bibitem[\protect\citeauthoryear{Floyd and Steinberg}{Floyd and
  Steinberg}{1976}]{floyd-steinberg}
{\sc Floyd, R.} {\sc and} {\sc Steinberg, L.} 1976.
\newblock An adaptive algorithm for spatial grey scale.
\newblock {\em Proc. of the Soc. of Info. Display\/}~{\em 17,\/}~1.

\bibitem[\protect\citeauthoryear{Ha\v{s}an, Fuchs, Matusik, Pfister, and
  Rusinkiewicz}{Ha\v{s}an et~al\mbox{.}}{2010}]{Hasan2010}
{\sc Ha\v{s}an, M.}, {\sc Fuchs, M.}, {\sc Matusik, W.}, {\sc Pfister, H.},
  {\sc and} {\sc Rusinkiewicz, S.} 2010.
\newblock Physical reproduction of materials with specified subsurface
  scattering.
\newblock {\em {ACM} TOG (Proc. SIGGRAPH)\/}~{\em 29,\/}~3.

\bibitem[\protect\citeauthoryear{Hergel and Lefebvre}{Hergel and
  Lefebvre}{2014}]{hergel2014}
{\sc Hergel, J.} {\sc and} {\sc Lefebvre, S.} 2014.
\newblock Clean color: Improving multi-filament {3D} prints.
\newblock {\em CGF (Proc. Eurographics)\/}~{\em 33,\/}~2, 469--478.

\bibitem[\protect\citeauthoryear{ICC}{ICC}{2010}]{ICC43}
{\sc ICC}. 2010.
\newblock {\em File Format for Color Profiles\/}, 4.3.0.0 ed.
\newblock http://www.color.org.

\bibitem[\protect\citeauthoryear{Johnson and Fairchild}{Johnson and
  Fairchild}{2003}]{JohnsonFairchild2003}
{\sc Johnson, G.} {\sc and} {\sc Fairchild, M.} 2003.
\newblock {A top down description of S-CIELAB and CIEDE2000}.
\newblock {\em Color Research and Application\/}~{\em 28,\/}~6, 425--435.

\bibitem[\protect\citeauthoryear{Lan, Dong, Pellacini, and Tong}{Lan
  et~al\mbox{.}}{2013}]{Lan2013}
{\sc Lan, Y.}, {\sc Dong, Y.}, {\sc Pellacini, F.}, {\sc and} {\sc Tong, X.}
  2013.
\newblock Bi-scale appearance fabrication.
\newblock {\em {ACM} TOG (Proc. SIGGRAPH)\/}~{\em 32,\/}~4.

\bibitem[\protect\citeauthoryear{Lau, Arce, and Gallagher}{Lau
  et~al\mbox{.}}{1999}]{LauArceGallagher1999}
{\sc Lau, D.}, {\sc Arce, G.}, {\sc and} {\sc Gallagher, N.} 1999.
\newblock Digital halftoning by means of green-noise masks.
\newblock {\em JOSA A\/}~{\em 16,\/}~7, 1575--1586.

\bibitem[\protect\citeauthoryear{Lau and Arce}{Lau and
  Arce}{2001}]{LauArce2001}
{\sc Lau, D.~L.} {\sc and} {\sc Arce, G.~R.} 2001.
\newblock {\em Modern digital halftoning}.
\newblock CRC Press.

\bibitem[\protect\citeauthoryear{Levin, Glasner, Xiong, Durand, Freeman,
  Matusik, and Zickler}{Levin et~al\mbox{.}}{2013}]{Levin_siggraph2013}
{\sc Levin, A.}, {\sc Glasner, D.}, {\sc Xiong, Y.}, {\sc Durand, F.}, {\sc
  Freeman, W.}, {\sc Matusik, W.}, {\sc and} {\sc Zickler, T.} 2013.
\newblock Fabricating {BRDF}s at high spatial resolution using wave optics.
\newblock {\em {ACM} TOG (Proc. SIGGRAPH)\/}~{\em 32,\/}~4.

\bibitem[\protect\citeauthoryear{Li and Allebach}{Li and
  Allebach}{2004}]{Li_tded_2004}
{\sc Li, P.} {\sc and} {\sc Allebach, J.} 2004.
\newblock Tone-dependent error diffusion.
\newblock {\em {IEEE} Trans. on Image Proc.\/}~{\em 13,\/}~2, 201--215.

\bibitem[\protect\citeauthoryear{Lou and Stucki}{Lou and
  Stucki}{1998}]{lou_stucki_1998}
{\sc Lou, Q.} {\sc and} {\sc Stucki, P.} 1998.
\newblock Fundamentals of {3D} halftoning.
\newblock {\em LNCS (Proc. Elect. Pub. and Art. Imag.)\/}~{\em 1375}, 224--239.

\bibitem[\protect\citeauthoryear{{MC}or Technologies}{{MC}or
  Technologies}{2014}]{Iris}
{\sc {MC}or Technologies}. 2014.
\newblock {MC}or {I}ris.
\newblock http://mcortechnologies.com/3d-printers/iris/.

\bibitem[\protect\citeauthoryear{Mitsa and Parker}{Mitsa and
  Parker}{1992}]{MitsaParker1992}
{\sc Mitsa, T.} {\sc and} {\sc Parker, K.} 1992.
\newblock Digital halftoning technique using a blue-noise mask.
\newblock {\em JOSA A\/}~{\em 9,\/}~11, 1920--1929.

\bibitem[\protect\citeauthoryear{Morovi{\v{c}}}{Morovi{\v{c}}}{2008}]{Morovic2008}
{\sc Morovi{\v{c}}, J.} 2008.
\newblock {\em {Color Gamut Mapping}}.
\newblock John Wiley \& Sons.

\bibitem[\protect\citeauthoryear{Mullen}{Mullen}{1985}]{Mullen1985}
{\sc Mullen, K.~T.} 1985.
\newblock {The contrast sensitivity of human colour vision to red-green and
  blue-yellow chromatic gratings.}
\newblock {\em The Jour. of Physiology\/}~{\em 359,\/}~1, 381.

\bibitem[\protect\citeauthoryear{Ostromoukhov}{Ostromoukhov}{2001}]{ostromoukhov_errdiff_2001}
{\sc Ostromoukhov, V.} 2001.
\newblock A simple and efficient error-diffusion algorithm.
\newblock In {\em Proc. SIGGRAPH}.

\bibitem[\protect\citeauthoryear{Pang, Qu, Wong, Cohen-Or, and Heng}{Pang
  et~al\mbox{.}}{2008}]{PangQuWongCohen-OrHeng2008}
{\sc Pang, W.-M.}, {\sc Qu, Y.}, {\sc Wong, T.-T.}, {\sc Cohen-Or, D.}, {\sc
  and} {\sc Heng, P.} 2008.
\newblock Structure-aware halftoning.
\newblock {\em {ACM} {TOG} (Proc. SIGGRAPH)\/}~{\em 27,\/}~3, 89.

\bibitem[\protect\citeauthoryear{Reiner, Carr, Mech, Stava, Dachsbacher, and
  Miller}{Reiner et~al\mbox{.}}{2014}]{reiner2014}
{\sc Reiner, T.}, {\sc Carr, N.}, {\sc Mech, R.}, {\sc Stava, O.}, {\sc
  Dachsbacher, C.}, {\sc and} {\sc Miller, G.} 2014.
\newblock Dual-color mixing for fused deposition modeling printers.
\newblock {\em CGF (Proc. Eurographics)\/}~{\em 33,\/}~2, 479--486.

\bibitem[\protect\citeauthoryear{Rogers}{Rogers}{1997}]{Rogers1997}
{\sc Rogers, G.~L.} 1997.
\newblock {Optical Dot Gain in a Halftone Print}.
\newblock {\em JIST\/}~{\em 41}, 643--656.

\bibitem[\protect\citeauthoryear{ScanLab and TurboSquid}{ScanLab and
  TurboSquid}{2013}]{Ruslan_TurboSquid}
{\sc ScanLab} {\sc and} {\sc TurboSquid}. 2013.
\newblock http://www.turbosquid.com/FullPreview/Index.cfm/ID/777450.

\bibitem[\protect\citeauthoryear{Schmidt, Grimm, and Wyvill}{Schmidt
  et~al\mbox{.}}{2006}]{schmidt_dexp_2006}
{\sc Schmidt, R.}, {\sc Grimm, C.}, {\sc and} {\sc Wyvill, B.} 2006.
\newblock Interactive decal compositing with discrete exponential maps.
\newblock {\em {ACM} TOG (Proc. SIGGRAPH)\/}~{\em 25,\/}~3, 605--613.

\bibitem[\protect\citeauthoryear{Stratasys}{Stratasys}{2014}]{Connex3}
{\sc Stratasys}. 2014.
\newblock Objet500 {Connex3}.
\newblock
  http://www.stratasys.com/3d-printers/production-series/connex3-systems.

\bibitem[\protect\citeauthoryear{S{\"u}sstrunk, Buckley, and
  Swen}{S{\"u}sstrunk et~al\mbox{.}}{1999}]{SuesstrunkBuckleySwen1999}
{\sc S{\"u}sstrunk, S.}, {\sc Buckley, R.}, {\sc and} {\sc Swen, S.} 1999.
\newblock {Standard RGB color spaces}.
\newblock In {\em IS\&T/SID, 7th CIC}. Scottsdale Ariz., 127--134.

\bibitem[\protect\citeauthoryear{Ten24}{Ten24}{2013}]{Head_Ten24}
{\sc Ten24}. 2013.
\newblock http://www.ten24.info/?p=1164.

\bibitem[\protect\citeauthoryear{TurboSquid}{TurboSquid}{2010}]{Apple_TurboSquid}
{\sc TurboSquid}. 2010.
\newblock http://www.turbosquid.com/3d-models/free-max-model-apple/549455.

\bibitem[\protect\citeauthoryear{Ulichney}{Ulichney}{1987}]{Ulichney1987}
{\sc Ulichney, R.} 1987.
\newblock {\em Digital Halftoning}.
\newblock The MIT Press.

\bibitem[\protect\citeauthoryear{{Van Nes} and Bouman}{{Van Nes} and
  Bouman}{1967}]{VanNesBouman1967}
{\sc {Van Nes}, F.~L.} {\sc and} {\sc Bouman, M.~A.} 1967.
\newblock Spatial modulation transfer in the human eye.
\newblock {\em JOSA\/}~{\em 57,\/}~3, 401--406.

\bibitem[\protect\citeauthoryear{Vidim\v{c}e, Wang, Ragan-Kelley, and
  Matusik}{Vidim\v{c}e et~al\mbox{.}}{2013}]{Vidimce_openfab_2013}
{\sc Vidim\v{c}e, K.}, {\sc Wang, S.-P.}, {\sc Ragan-Kelley, J.}, {\sc and}
  {\sc Matusik, W.} 2013.
\newblock Openfab: A programmable pipeline for multi-material fabrication.
\newblock {\em {ACM} TOG (Proc. SIGGRAPH)\/}~{\em 32,\/}~4.

\bibitem[\protect\citeauthoryear{Vora and Trussell}{Vora and
  Trussell}{1993}]{VoraTrussell1993}
{\sc Vora, P.~L.} {\sc and} {\sc Trussell, H.~J.} 1993.
\newblock Measure of goodness of a set of color-scanning filters.
\newblock {\em JOSA A\/}~{\em 10}, 1499--1508.

\bibitem[\protect\citeauthoryear{Wyble and Berns}{Wyble and
  Berns}{2000}]{WybleBerns2000}
{\sc Wyble, D.~R.} {\sc and} {\sc Berns, R.~S.} 2000.
\newblock {A Critical Review of Spectral Models Applied to Binary Color
  Printing}.
\newblock {\em Color Research and Application\/}~{\em 25,\/}~1, 4--19.

\bibitem[\protect\citeauthoryear{Wyszecki and Stiles}{Wyszecki and
  Stiles}{2000}]{WyszeckiStiles2000}
{\sc Wyszecki, G.} {\sc and} {\sc Stiles, W.} 2000.
\newblock {\em Color Science: Concepts and Methods, Quantitative Data and
  Formulae\/}, 2 ed.
\newblock John Wiley \& Sons, inc.

\bibitem[\protect\citeauthoryear{Zhang and Wandell}{Zhang and
  Wandell}{1996}]{ZhangWandell1996}
{\sc Zhang, X.} {\sc and} {\sc Wandell, B.~A.} 1996.
\newblock {A spatial extension of CIELAB for digital color image reproduction}.
\newblock {\em Society for Information Display Symposium Technical
  Digest\/}~{\em 27}, 731--734.

\bibitem[\protect\citeauthoryear{Zhou and Fang}{Zhou and
  Fang}{2003}]{ZhouFang_varcoeff_threshmod_2003}
{\sc Zhou, B.} {\sc and} {\sc Fang, X.} 2003.
\newblock Improving mid-tone quality of variable-coefficient error diffusion
  using threshold modulation.
\newblock {\em {ACM} TOG (Proc. SIGGRAPH)\/}~{\em 22,\/}~3, 437--444.

\end{thebibliography}

\section{Supplemental Material}

\subsection{Radiative transfer}
Given an arrangement of multiple non-fluorescent printing materials with similar refractive indexes within a shape $\shapeS$, light propagation within this shape can be described by the steady-state radiative transfer equation~\cite{Chandrasekhar1960}
\begin{align} 
 \frac{\mathrm{d}I_{\lambda}(x,\Omega)}{\mathrm{d}x}  = &-a_{\lambda}(x)I_{\lambda}(x,\Omega) \label{eq:RadiativeTransfer} \\
&- s_{\lambda}(x)\Big(I_{\lambda}(x,\Omega) - \oint_{\mathbb{S}^2} I_{\lambda}(x,\Omega')p_{\lambda}(\Omega',\Omega)\mathrm{d}\Omega'\Big) \nonumber
\end{align} 
where $I_{\lambda}(x,\Omega)$ is the radiant intensity at location $x \in \shapeS$ propagating in direction $\Omega$ for wavelength $\lambda \in [380,730]$ nm (visible wavelength range), $a_{\lambda}(x) \geq 0$ is the spectral absorption coefficient, $s_{\lambda}(x) \geq 0$ is the spectral scattering coefficient, $\mathbb{S}^2$ is the sphere and $p_{\lambda}(\Omega', \Omega)$ is the scattering function satisfying $\oint_{\mathbb{S}^2}\oint_{\mathbb{S}^2} p_{\lambda}(\Omega',\Omega)\mathrm{d}\Omega'\mathrm{d}\Omega = 1$. 
Solving the radiative transfer equation with an appropriate boundary condition at the surface gives us the radiation that is diffusely emitted from the 
shape. Adding the radiation directly reflected from the surface due to Fresnel reflection, gives us the total radiation emitted from the object's surface.

Eq. \eqref{eq:RadiativeTransfer} shows that the intensity of radiant energy traveling through the material is attenuated by absorption (note that $a_{\lambda}(x)$ depends on the wavelength but is independent of the traveling direction) and is redistributed by scattering (scaled by $s_{\lambda}(x)$). Thus, a fraction of light entering the print at one location may be emitted from the surface at a different location due to scattering. If light travels through different materials its spectral power distribution is modulated by each material's absorption coefficients and the path length within this material.

\subsection{Gamut Volume Visualizations}
\begin{figure*}[b]
\centering
\parbox{0.25\textwidth}{
\centering
\includegraphics[width=0.25\textwidth]{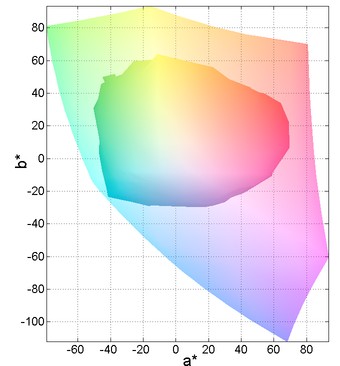}\\0.269
}\hfill
\parbox{0.25\textwidth}{
\centering
\includegraphics[width=0.25\textwidth]{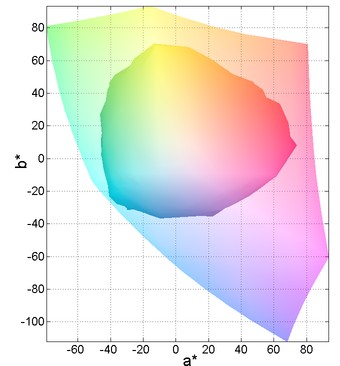}\\0.322
}\hfill
\parbox{0.25\textwidth}{
\centering
\includegraphics[width=0.25\textwidth]{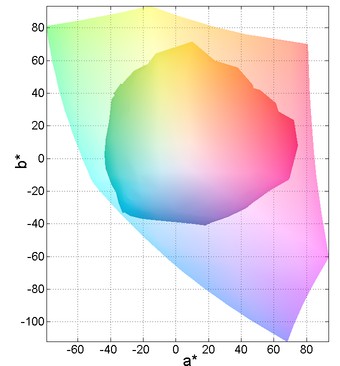}\\0.372
}\hfill
\parbox{0.25\textwidth}{
\centering
\includegraphics[width=0.25\textwidth]{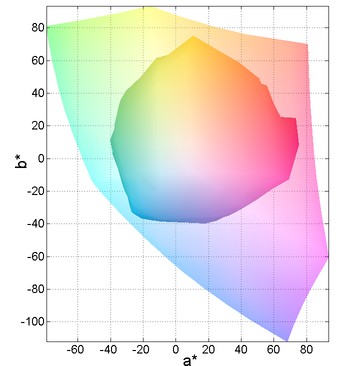}\\0.409
}
\caption{Increasing number of layers results in increasing gamut volume. From left to right: 3, 6, 12 and 18 layers. The gamut volume is indicated below each image as a fraction of the volume of sRGB gamut shown transparent.}
\label{fig:layer_gamut_ab}
\end{figure*}

Figure \ref{fig:layer_gamut_ab} shows the gamut volumes from the paper projected onto the $a^{*} b^{*}$ plane to provide another perspective on how the gamut volume changes as layers are added.

\subsection{Visual Experiment Setup}
\begin{figure*}
\includegraphics[width=\textwidth]{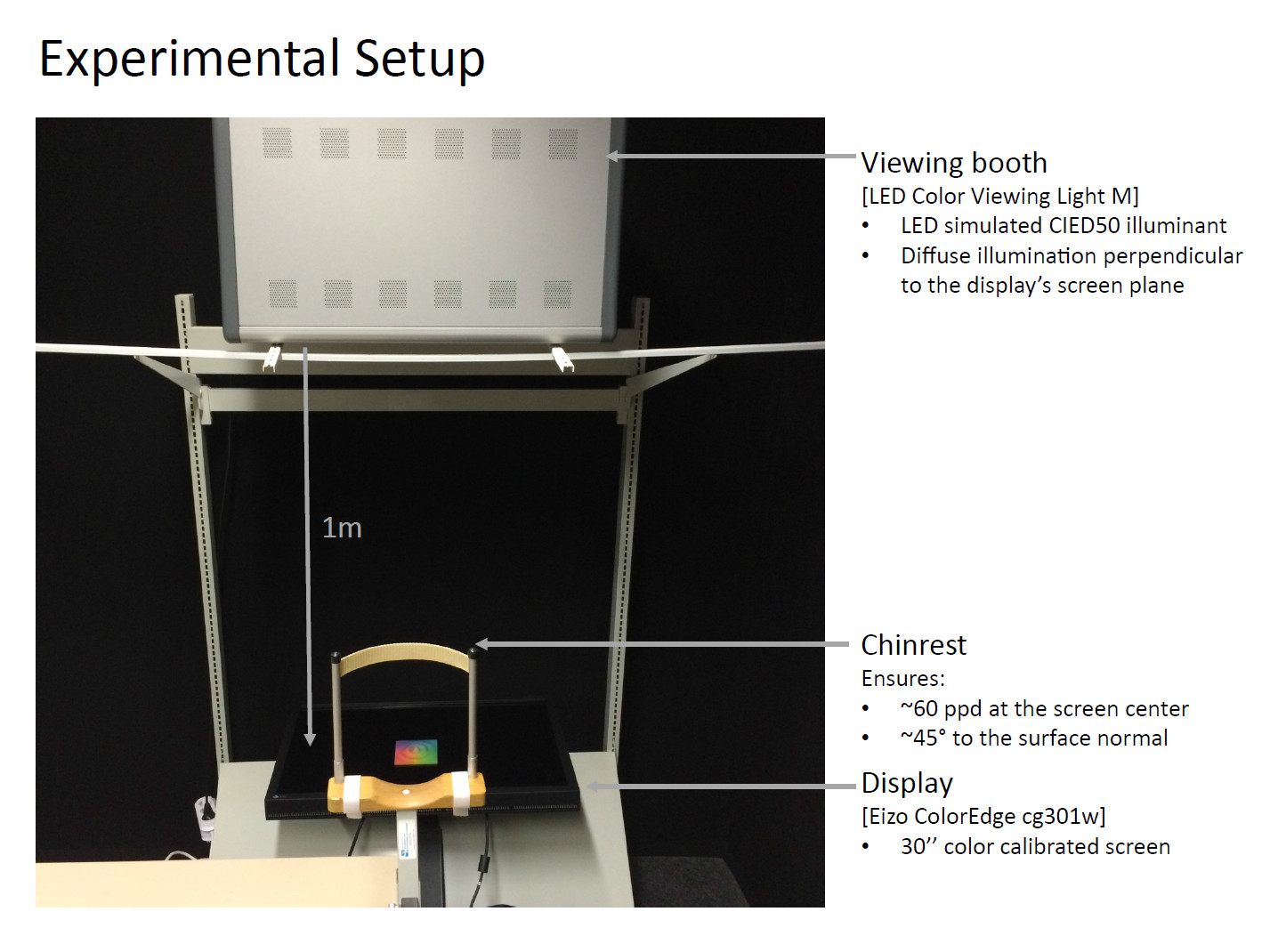}
\caption{Physical setup for our visual experiment.}
\label{fig:experiment_setup}
\end{figure*}

Figure \ref{fig:experiment_setup} shows the physical setup we used to conduct the visual experiment. Subject placed their chin on the chinrest and viewed the printed object next to a rendering of it on a color-calibrated display. The color characterization of the display was performed from the chinrest position considering the viewing booth illumination. In this way, tristimulus values from objects placed on the display and illuminated by the viewing booth could be reproduced on the screen if viewed from the chinrest position.

\end{document}